\documentclass[12pt,a4paper]{article}

\newif\ifsubmit
\submitfalse

\usepackage{authblk}  

\usepackage{algorithm}
\usepackage{algpseudocode}
\usepackage{tikz}
\usepackage{booktabs} 

\usepackage{here}
\usepackage{amsmath}
\usepackage[hyphens]{url}
\usepackage{eqnarray}
\usepackage{multirow}
\usepackage{cite}
\usepackage{float}
\usepackage{comment}
\usepackage{subcaption}
\usepackage[flushleft]{threeparttable}
\usepackage{adjustbox}
\usepackage{lscape}

\newif\ifdraft
\draftfalse 

\newif\ifpubdraft
\pubdraftfalse

\newif\ifrevdraft
\revdraftfalse

\newif\ifthesisdraft
\thesisdraftfalse

\newif\ifthesisdraftTwo
\thesisdraftTwofalse

\ifpubdraft
        \usepackage{ulem}
        \newcommand{\dkawai}[1]{\textcolor{red}{#1}} 
        \newcommand{\dkawaiErase}[1]{\textcolor{red}{\sout{#1}}} 
\else
        \newcommand{\dkawai}[1]{#1}
        \newcommand{\dkawaiErase}[1]{}
\fi

\ifdraft
        \newcommand{\daisuke}[1]{\textcolor{red}{[[Daisuke: #1]]}}
        \newcommand{\nicolasc}[1]{\textcolor{brown}{[[Nicolas: #1]]}}
        \newcommand{\acv}[1]{\textcolor{teal}{[[Alejandro: #1]]}}
        \newcommand{\ks}[1]{\textcolor{olive}{[[Kyle: #1]]}}
\else
        \newcommand{\daisuke}[1]{}
        \newcommand{\nicolasc}[1]{}
        \newcommand{\acv}[1]{}
        \newcommand{\ks}[1]{}
\fi

\ifrevdraft
        \usepackage{ulem}
\else
        
\fi

\ifthesisdraft
        \usepackage{ulem}
        \newcommand{\dkAdd}[1]{\textcolor{red}{#1}} 
\else
        \newcommand{\dkAdd}[1]{#1}
\fi

\ifthesisdraftTwo
        \usepackage{ulem}
        \newcommand{\dkAddTwo}[1]{\textcolor{red}{#1}} 
        \newcommand{\dkEraseTwo}[1]{\textcolor{red}{\sout{#1}}} 
\else
        \newcommand{\dkAddTwo}[1]{#1}
        \newcommand{\dkEraseTwo}[1]{} 
\fi        

\usepackage[square,numbers]{natbib}
\begin{document}

\title{Anatomy of a Digital Bubble: Lessons Learned from the NFT and Metaverse Frenzy}


\newif\ifanon
\anonfalse

\ifanon
  \author{Anonymous submission}
\else
  \author[1]{Daisuke Kawai\thanks{dkawai@alumni.cmu.edu}}
  \author[2]{Kyle Soska\thanks{ksoska@alumni.cmu.edu}}
  \author[1]{Bryan Routledge\thanks{routledge@cmu.edu}}
  \author[1]{\\Ariel Zetlin-Jones\thanks{azj@andrew.cmu.edu}}
  \author[1]{Nicolas Christin\thanks{nicolasc@andrew.cmu.edu}}
  \affil[1]{Carnegie Mellon University, Pittsburgh, Pennsylvania, USA}
  \affil[2]{Independent, New York, New York, USA}
\fi


\maketitle

\begin{abstract}
  In the past few years, ``metaverse'' and ``non-fungible tokens (NFT)''
have become buzzwords, and the prices of related assets have exhibited
large fluctuations. Are those characteristic of a speculative bubble? In
this paper, we attempt to answer this question, and better understand
the underlying economic dynamics. Specifically, we look at Decentraland,
a virtual world platform where land parcels are sold as NFT collections.
We find that initially, land prices followed traditional real estate
pricing models -- in particular, value decreased with distance from
the most desirable areas -- suggesting Decentraland behaved much like
a virtual city. However, these real estate pricing models stopped
applying when both the metaverse and NFTs gained increased popular
attention and enthusiasm in 2021, suggesting a new driving force for
the underlying asset prices. At that time, following a substantial
rise in NFT market values, short-term holders of multiple parcels
began to take major selling positions in the Decentraland market,
which hints that, rather than building a metaverse community, early
Decentraland investors preferred to cash out when land valuations became
inflated. Our analysis also shows that while the majority of buyers are
new entrants to the market (many of whom joined during the bubble),
liquidity (i.e., parcels) was mostly provided by early adopters selling,
which caused stark differences in monetary gains. Early adopters made
money -- more than 10,000 USD on average per parcel sold -- but users
who joined later typically made no profit or even incurred losses in
the order of 1,000 USD per parcel. Unlike established markets such as
financial and real estate markets, newly emergent digital marketplaces
are mostly self-regulated. As a result, the significant financial
risks we identify indicate a strong need for establishing appropriate
standards of business conduct and improving user awareness.

\end{abstract}



\section{Introduction}
\label{sec:dl_intro}
The word ``\textit{metaverse}'' was originally coined in science fiction
in 1992 \cite{Joshua_2017}, to denote an extension of
the Internet into a single virtual world. With advances in computer
graphics and virtual reality hardware, we have seen increasingly
realistic attempts to build a metaverse -- or at least a number of
virtual worlds -- over the past couple of decades, with platforms
such as Second Life becoming quite popular. 
Tellingly, in a bid that virtual worlds were about to harness enormous
economic potential, in October 2021, Facebook rebranded itself as
``Meta,'' hoping to become one of the main building blocks of this
metaverse \cite{Meta_2021}.

Until recently, most virtual worlds (e.g., Second Life, World
of Warcraft, etc.) were centralized, and operated by a single company.
Although second-hand markets exist, in traditional, centralized
virtual worlds, users willing to purchase wearables for instance 
to decorate their virtual-world avatars or acquire virtual land,
usually have to buy them from the hosting company.

On the other hand, newer projects, such as
Decentraland\footnote{\url{https://decentraland.org/}.} or The
Sandbox,\footnote{\url{https://www.sandbox.game/en/}.} have
leveraged recent advances in decentralization technologies, e.g.,
``blockchains'' and cryptocurrencies 
that instantiate bankless peer-to-peer payment
systems. These newer projects rely on a blockchain 
to host vital records of their underlying virtual worlds.
Further, governance, instead of being centralized, is handled by 
decentralized anonymous organizations (DAOs). Roughly speaking, DAOs
attempt to implement a democratic process where governance tokens can
be purchased by the general public. Similar to holding company
shares, votes 
are typically weighted by the amount of
governance tokens one holds \cite{Junque_Zhang_2023}. 
Likewise, decentralized virtual world
items (e.g., wearables, land, and so forth) are often instantiated
as ``non-fungible tokens'' (NFTs), which are also supported by
(distributed) blockchain infrastructure rather than a hosting company.
Importantly, \emph{both} governance tokens and NFTs 
can be traded \textit{outside} the virtual world platform, e.g.,
on cryptocurrency exchanges, or third-party NFT marketplaces such as
OpenSea.\footnote{\url{https://opensea.io}.}

Decentralized virtual worlds appear to 
have capitalized on the surge in popularity of both the metaverse and NFTs.
Concretely, in late 2021, the renewed interest in the metaverse following Facebook's announcement, combined with the simultaneous sudden and 
substantial rise in cryptocurrency prices, caused a massive increase in 
Decentraland virtual land (parcel) prices, and in the valuation of its 
governance token called MANA. The median sales price of land in 
Decentraland rose from less than 1,000~USD in 2020 to roughly 
15,000~USD in Q4 of 2021---only to plummet back to 5,000~USD in the 
subsequent half year.
By the end of 2023, prices had gone back down to approximately 
1,000~USD, and, as we will see, activity was minimal. At the time of writing, 
parcels go for approximately 200 USD. 

Such volatility shows the hallmarks of an economic bubble. 
Was that indeed the case?
How did the situation affect the wealth of 
all investors participating in Decentraland---both buyers and sellers?
Were 
certain groups of users disproportionately and/or adversely affected?
If so, could regulation help ``level the playing field?'' 
Our main
contribution is addressing these questions based on a quantitative analysis
of online conversations on Reddit and of Ethereum blockchain records
about Decentraland.

We argue this is an important contribution to the study of the economics of digital markets.
Economic literature has documented numerous economic bubbles across various markets and countries \cite{Ofek_Richardson_2003, Penasse_Renneboog_2022, Liu_Zhang_Zhao_2015}. 
However, these markets are typically geared towards accredited investors, high-income individuals, or otherwise limited segments of the population, resulting in limited exposure for the general public.
On the other hand, despite their complex technological underpinnings, the markets for NFTs and metaverse-related asssets are directly marketed to the ordinary investors---including children \cite{SJVB2023}. 
Faverio~and~Sidoti~\cite{Faverio_Sidoti_2023} shows that 17\% of U.S. citizens, particularly young adults (e.g., 41\% of young men), hold or have held cryptocurrencies. 
Given the similarities between cryptocurrencies and NFTs (common blockchain
lineage, novel technology, etc.), 
we conjecture that cryptocurrency 
investors are also likely to bet on NFTs.
Therefore, we argue that 
documenting the impact of the NFT and metaverse economic bubble 
could, more generally, help understand the effects of 
modern ``digital bubbles,'' which 
impact a segment of the population potentially broader than ever before. 

To do so, 
we first examine relevant Reddit conversations to confirm that the 
Decentraland platform is representative of the metaverse and NFT ecosystem. 
Through topic modeling, we show that roughly 27\% of the posts 
discussing Decentraland are about metaverse and NFT; the second-largest 
topic discussed relates to investments in metaverse projects (not limited 
to  Decentraland).
Even more interestingly, online activity (i.e., number of posts) 
related to these topics closely correlates with the trading volume of 
Decentraland land parcels.
Furthermore, we observe that as land parcel prices are declining, violations
of Reddit terms-of-use\footnote{https://www.redditinc.com/policies/} in the 
relevant forums increase. This hints that users are getting frustrated 
by the economic downturn, and  are forgoing civil discourse as the general 
interest from the broader online community is waning.

Second, we examine whether traditional (physical) real estate pricing
models apply to Decentraland land prices to understand if at some time
these prices reflected any traditional notions of the fundamental
value of real estate. Decentraland uses a finite Euclidean topology to
subdivide the entire world into parcels, each of which is paired to a
unique NFT. Similar to holding a property title in the physical world,
the owner of that NFT owns the parcel, and, accordingly, can modify
its contents (e.g., build something), start a business, or sell it.
Different from the physical world, though, Decentraland allows users
to instantly teleport anywhere within the virtual world. The question
then becomes whether parcel prices follow traditional real-estate price
patterns, with prices decreasing as a function of distance from a
desirable location. This would indicate a property characteristic of a 
``true'' virtual city. 

We show that in the early period when prices were low, parcel prices do in fact appear to follow this class of traditional real-estate models quite closely.
But then, during the period of the rapid rise in land prices, location
no longer appears to matter much--investors seem to purchase plots like
they purchase other NFTs (e.g., art or collectibles), presumably for
speculative purposes. In other words, land prices begin to deviate from
the traditional real-estate models, suggesting a bubble-like phenomenon.

Our analysis of market demographics further shows that two distinct groups
played major roles during the period we study. New entrants (i.e., those
who started trading during the bubble), most of whom appear to
be individual investors, were dominantly net-buyer groups (i.e., they
bought more than they sold). On the other hand, early adopters provided
liquidity (i.e., parcels) to the market and were mostly
selling. This caused stark differences in monetary profits: Early
adopters reaped 15,000 USD worth of profit per parcel on average. On
the other hand, new entrants could rarely profit from the sales -- in
general, they lost in the order of 1,000 USD per parcel. In
short, the time at which prices were high 
led to a disproportionate flow of money
from new participants to early adopters.

Digital item marketplaces fostered by the metaverse and NFT
are currently subject to lax or ill-defined regulations \cite{EU_2023,
Kumar_McLaughlim_Xie_Nicolet-Serra_Muller_Rigg, SEC_NFT}. However, our results
highlight significant financial risk to new market entrants, who may
have been lured by the promise of riches, when the situation was
markedly unstable. These new market entrants may not have had enough
information to make rational or well-informed decisions. 
This result echoes the negative impact of the frenzy surrounding blockchain 
technology \cite{Cheng_Franco_Jiang_Lin_2019} and emphasizes the 
need for establishing appropriate standards of business conduct and 
improving user awareness, for example, through education or mandatory 
risk disclosure.
 
\section{Related work and background}
\label{sec:dl_literature}

\paragraph{Economic bubbles}
An economic bubble is broadly 
defined as a situation where 
a substantial rise in asset prices causes them to 
greatly exceed their fundamental value.
Bubbles often follow inflated speculative trades and eventually result in 
significant investor losses \cite{Ofek_Richardson_2003, Hong_Stein_2007}.
For cryptocurrencies specifically, Cheng~et~al.~\cite{Cheng_Franco_Jiang_Lin_2019} 
shows that investments in blockchain companies during the 2017 
initial coin offering (ICO) boom led to negative performance, 
and presumably to considerable investor losses.
Due to their negative societal 
impact, the study of economic bubbles and of their 
root causes 
has been a focal interest of economics and monetary policy \cite{Asriyan_Fornaro_Martin_Ventura_2021, Gertler_Kiyotaki_Prestipino_2020}.
A frequently cited cause of economic bubbles is that investors 
make biased decisions based on an extrapolation of past returns 
\cite{Case_Shiller_Thompson_2012, Cutler_Poterba_Summers_1990, De_Long_Shleifer_Summers_Waldmann_1990, Hong_Stein_1999}.
As early as 1977, Minsky~\cite{Minsky_1977} claims that investors' subjective 
extrapolation of past returns amplifies economic cycles (``financial 
instability hypothesis'').
Aliber~et~al.~\cite{Aliber_Kindleberger_McCauley_2023} argue that economic bubbles 
evolve upon the arrival of good news leading speculative investors to overvalue asset  prices.
As a case in point, 
Cooper~et~al.~\cite{Cooper_Dimitrov_Rau_2001} shows that,
during the ``dotcom'' bubble of 
the 1990s,
merely changing 
company names to include references to the Internet 
had a substantial positive effect on the corresponding stock prices.

Recent literature also proposes extrapolation-based models of economic bubbles \cite{Jin_Sui_2022, Barberis_Greenwood_Jin_Shleifer_2015, Adam_Marcet_Beutel_2017, Glaeser_Nathanson_2017, Cassella_Gulen_2018, DeFusco_Nathanson_Zwick_2017, Barberis_Greenwood_Jin_Shleifer_2018}.
For instance, Barberis~et~al.~\cite{Barberis_Greenwood_Jin_Shleifer_2018}'s model features a mixture of rational and extrapolating investors and describes the mechanism by which the latter dominate the market when good news is published, leading to an economic bubble.
A remarkable feature of their model is that it explains the inflated trading volume during the bubble period.

Another thesis is that bubbles are caused by 
investors' underestimation of downside risk.
Gennaioli~et~al.~\cite{Gennaioli_Shleifer_Vishny_2012}'s model  
show that unconscious exclusion of bad scenarios causes economic crises. 
Specifically, market participants ignoring bad scenarios leads to excessive 
issuance of risky assets. Unfortunately, as soon as market participants 
become aware of the downside risks, they abruptly withdraw funds, leading 
to steep declines in underlying asset value.

Bordalo~et~al.~\cite{Bordalo_Gennaioli_Shleifer_2018} propose ``diagnostic expectation''
as a mechanism to reconcile investors' extrapolation and their negligence of downside risks.
Their derivation stems from ``representative heuristics'' in the seminal work of Kahneman and Tversky~\cite{Kahneman_Tversky_1979}.
Diagnostic expectation posits that investors 
weigh recent news too heavily in extrapolating from 
past returns, leading to overly optimistic (resp. pessimistic) forecasts after good (resp. bad) news.
Maxted~\cite{Maxted_2023} incorporates diagnostic expectation to a real business cycle model 
and shows that optimistic (resp. pessimistic) outlooks cause increasing (resp. decreasing) asset prices, and that prices will revert to a rational level as optimism (resp. pessimism) fades away.

Diagnostic expectation may apply to the digital bubble we study. Indeed, 
media exposure of NFTs and of the metaverse abruptly surged in 2021, with notable events like the venerable Christie auction house 
holding an NFT auction, and Facebook's rebranding itself as Meta \cite{Christies_Beeple, Meta_2021}.
Investors may have expected a bright future for NFTs and for the metaverse, driven by the influx of news related to these concepts.

\paragraph{Metaverse and NFT}
While the metaverse is now being featured as an extension of our current digital experience \cite{Lee_Braud_Zhou_Wang_Xu_Lin_Kumar_Bermejo_Hui_2021, Mystakidis_2022}, Linden Lab launched Second Life (SL) as an online virtual world that accommodates massive number of players as early as 2003 \cite{Boellstorff_2015}.
A feature that purportedly contributed to its popularity is the economy in which users could profit by trading their generated items or virtual real estate \cite{Shelton_2010}.
While Messinger~et~al.~\cite{Messinger_Stroulia_Lyons_Bone_Niu_Smirnov_Perelgut_2009} suggest that making a profit is not the primary motivation for most users to play SL,
Bakshy~et~al.~\cite{Bakshy_Simmons_Huffaker_Cheng_Adamic_2010} documented the existence of an active market in SL and found that sellers employ social ties in the virtual world to increase their revenue.
As far as adverse effects, Stokes~\cite{Stokes_2012} points out the risk of money laundering facilitated by SL.

Second Life's market structure resembles recent attempts to instantiate a Metaverse.
On these platforms, users can profit by trading their created items and real estate published in the virtual worlds.
A striking difference from earlier attempts, though, is that they employ blockchain as a core data layer where the state of the world, such as land or item custody, is maintained.
For example, Decentraland has employed the Ethereum blockchain as a ledger for in-world items from its inception \cite{decentraland_wp}.

Public blockchains were originally introduced to implement a secure and transparent ledger system 
without relying on trusted third parties \cite{Nakamoto_2008}. 
However, this technology 
is now supporting a large volume of mostly speculative cryptocurrency trading \cite{Baur_Hong_Lee_2018, Soska_Dong_Khodaverdian_Zetlin-Jones_Routledge_Christin_2021, dkawai_2023}.
Besides cryptocurrencies, blockchains are also used to support non-fungible tokens (NFTs) \cite{Nadini_Alessandretti_Di_Giacinto_Martino_Aiello_Baronchelli_2021}.
Smart contracts, scripts to automate transactions, make secure and transparent tracking of each token's ownership possible, by following established NFT standards \cite{Entriken_Shirley_Evans_Sachs_2018}.
In particular, even though all transactions are executed on publicly available ledger systems,
NFT ownership is resistant to tampering, without relying on trusted third-parties.
Hence, NFTs have been prominently touted as a technology that facilitates the trade of creative works in a secure and transparent manner.
Another desirable feature of NFTs is that, being supported by decentralized blockchains, they are platform-independent. In particular, they are not tied to specific seller marketplaces: anybody can buy and sell them. 
As a result, metaverse platforms that use NFTs to 
represent 
in-world items allow these items to be traded outside the platform.

Our hope for this work is to complement the literature that has evidenced the possible negative impact of hype cycles surrounding blockchain technology 
\cite{Cheng_Franco_Jiang_Lin_2019, Junque_Zhang_2023}, by providing 
a quantitative analysis of the economic dynamics of the NFT market for 
metaverse-related items. More broadly, we hope to contribute to the 
pressing agenda of better understanding emerging ``Web3.0'' businesses 
\cite{Biais_Capponi_Cong_Gaur_Giesecke_2023}.
 
\section{Dataset}
\label{sec:dl_dataset}
This section briefly describes Decentraland first and introduces our dataset.
Our measurements and analysis are based on three data sources: (1) Ethereum
blockchain records, (2) Decentraland public APIs, and (3) 
Reddit posts.

\subsection{Decentraland dataset}
Decentraland is an online platform developed since
2015. Public interest in Decentraland grew in 2017 when developers
held an initial coin offering (ICO) of their \textit{MANA}
governance token along with the first land auction. In the auction,
participants had the option to bid on parcels of land or stake
their MANA toward community-initiated districts.\footnote{See
\url{https://medium.com/decentraland/the-terraform-event-is-around-the-corner-bdda6a2d4367} and
\url{https://github.com/decentraland/districts}.} Following a second
land auction in 2018 and the introduction of developer tools, the
official Decentraland client was launched in February 2020. At the time
of writing Decentraland has published roughly 93,000 parcels of land as
an NFT collection called \textit{LAND}.

Landowners can bundle up contiguous parcels they own into a new NFT token
called \textit{Estate (EST)}, host their business or events on their
parcels or estates, and list the parcels they own on
(third-party) NFT marketplaces for sale.

Similar to real estate transactions, buyers can either buy parcels
at the listed price, or bid a different price. They can also bid for
unlisted parcels and directly negotiate with landowners.

Two different LAND NFT tokens only differ in the 
address (formally, a dimensionless integer) they represent
in Decentraland. 
In particular, unlike many other NFT
collections, there is no specific artistic value or artificial scarcity
for a given LAND token. As such, we should be able to test whether
real-estate pricing models and economic theories apply to these virtual
parcels.

\paragraph{Ethereum blockchain records}
Ether (ETH), the utility token for the Ethereum blockchain, has the
second-largest market capitalization (after Bitcoin) at the time of
writing.\footnote{\url{https://coinmarketcap.com/}}
Ethereum's main feature is the ability to deploy smart contracts.
In a nutshell, smart contracts are programs independently
computed by all participants in the (Ethereum) network, who then
collectively reach a consensus on the program output. This allows 
computations to be extremely robust to attacks.
Decentraland employs smart contracts as a critical infrastructure. Because all
Ethereum transactions, including all interactions with smart contracts, 
are recorded in a public ledger (the Ethereum blockchain), we can
readily explore details of any Decentraland transaction, such as
the contract functions that were called, the parties to the
transaction, and any fees they might have incurred.

We collect all transactions interacting with the smart contracts listed on the Decentraland project page\footnote{https://github.com/decentraland/contracts} and third-party marketplaces where LAND and EST collections are listed until Dec.~31, 2023.
While some user-created NFTs available on Decentraland, such as wearables, are hosted on the Polygon blockchain, 
all transactions on LAND and EST collections are hosted on Ethereum,\footnote{https://docs.decentraland.org/player/market/marketplace/} so we focus on Ethereum. 
We validate our dataset by comparing it to Etherscan trade records.\footnote{\url{https://etherscan.io/token/0xf87e31492faf9a91b02ee0deaad50d51d56d5d4d\#tokenNFTTrade}}

Our dataset comprises four categories of transactions: (I) land purchases on the Decentraland marketplace, (II) land purchases on third-party marketplaces, (III) land auctions, and (IV) others (e.g., estate creation and dissolution).
We store all information, such as the breakdown of sales prices per parcel, the creation/dissolution of estates, and ownership transfers for LAND token sales (category (I)) and for all transactions in category (IV). 
However, our data collection faces a limitation for EST token sales in category (I) and a subset of (II) and (III).
Specifically, 
Decentraland held its first land auction off-chain, and at the time
only relied on the Ethereum blockchain to collect deposits to participate 
in 
the auction, to refund deposits after the auction, and to transfer 
parcels to auction winners. Therefore, we cannot determine the exact
price of each parcel if an auction participant won more than one parcel. 
Subsequent land auctions, 
bulk sales on third-party marketplaces, and estate
sales share the same price ambiguity, i.e., we cannot determine the cost
per parcel when somebody purchases more than one parcel.
Bulk sales on third-party marketplaces may in particular include NFT items not related to Decentraland.

\paragraph{Decentraland APIs}
Decentraland provides public APIs to collect parcel data,
including parcel property information and ownership transfer history, 
sorted chronologically.\footnote{\url{https://github.com/decentraland/LAND-permissions-graph}; \url{https://github.com/decentraland/atlas-server}.}
Every LAND token, including those bundled as EST tokens, has one or more
property tags attached to it to denote the type of property the parcel
refers to: 1) \textit{personally-owned lands} can be bought and sold by
users; 2) \textit{roads}, and 3) \textit{plazas}, on the other hand, are
controlled by the Decentraland development team. (In particular, the Genesis plaza, where
users spawn from, is in the center of the Decentraland map.) Finally, 4)
\textit{districts} denote a set of parcels developed by user communities
based on public proposals.
District founders have to obtain approval from the Decentraland DAO
community to launch their district. There are currently 39~districts on
the platform. A further 17~districts were dissolved in 2019.\footnote{\url{https://nftplazas.com/decentraland-districts/}}

\begin{figure}[htbp]
    \centering
    \includegraphics[width=0.50\columnwidth]{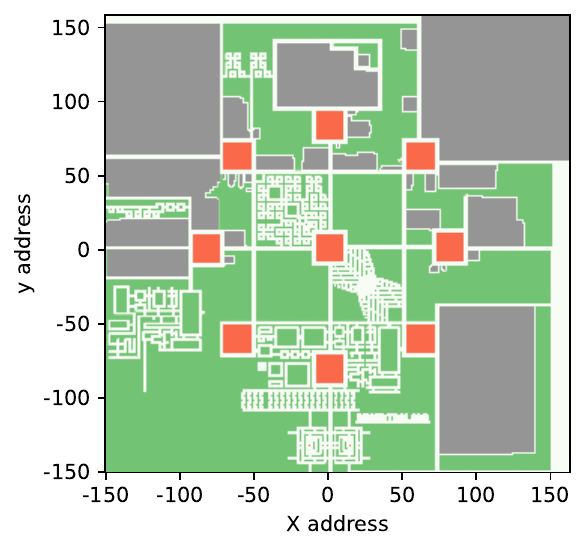}
    \caption{Map of Decentraland color-coded with parcel types: district (grey), plaza (orange), ownable parcels (green), and roads (white).}
    \label{fig:decentraland_map}
\end{figure}
Figure~\ref{fig:decentraland_map} shows the Decentraland map at the time of writing.
We collect all the ownership transfer records, estate creation and dissolution records, and land property information until Dec.~31, 2023 from the APIs.
Due to limitations on available property information, we cannot track
back the addresses of districts dissolved in 2019.

Additionally, Decentraland provides a dataset containing the history of building construction on parcels.
It has a comprehensive record of all buildings on parcels uploaded to the metaverse world.
An entry for the dataset includes all the details about the construction, such as the uploaded date, its contents, and builder metadata.

\subsection{Reddit dataset}
\begin{figure}[htbp]
    \centering
    \includegraphics[width=0.85\textwidth]{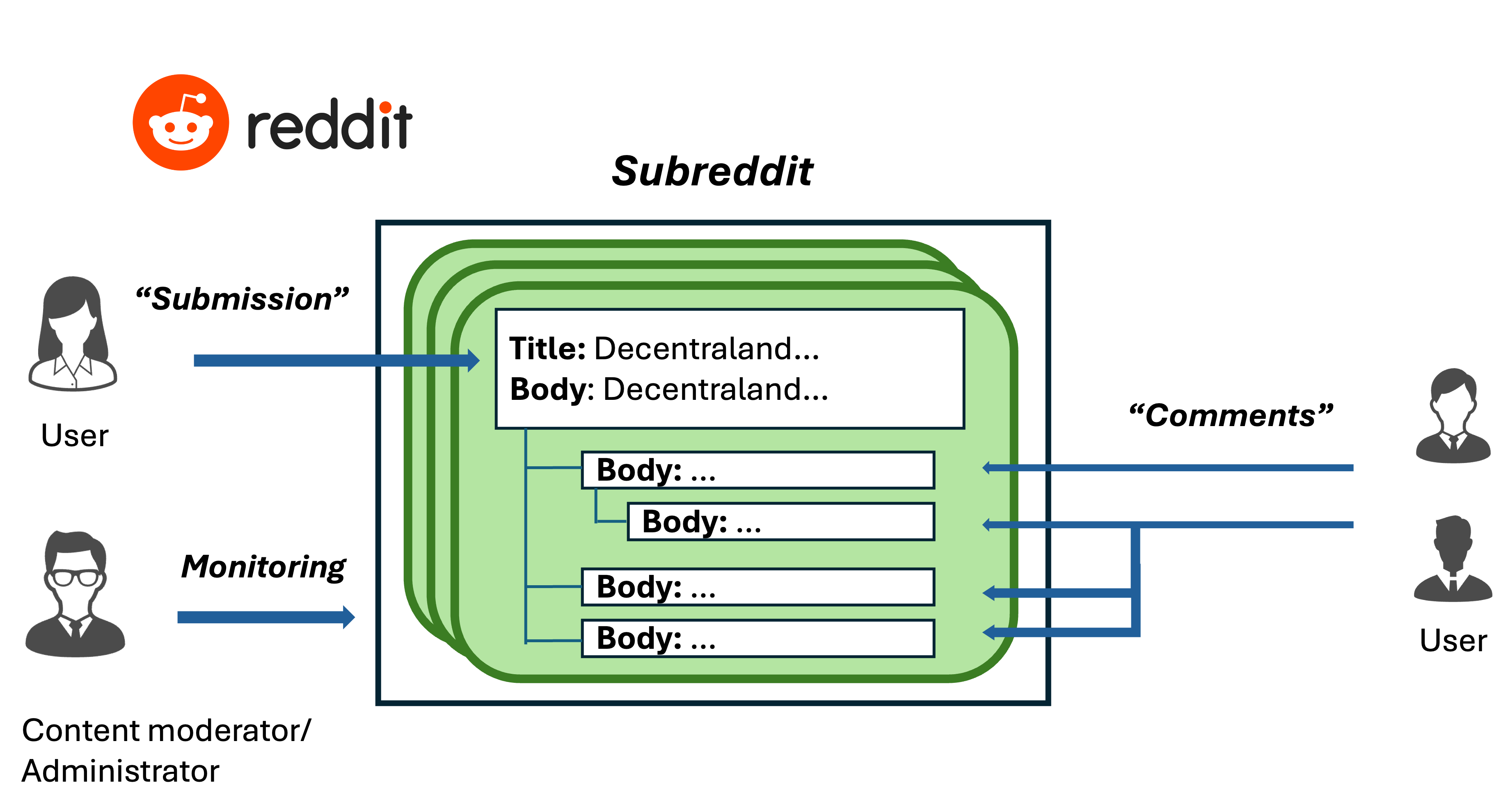}
    \caption{
    Reddit conversations.
    A user posts a ``submission'' under a relevant subreddit, and other users subsequently respond with ``comments.''
    Content moderators and administrators monitor the content posted on subreddits.
    }
    \label{fig:reddit_diagram}
\end{figure}

Reddit is a major social media platform mainly used by young adults \cite{Auxier_Anderson_2021}. 
Reddit users can post in forums dedicated to specific topics (e.g., relationship
advice or sports teams) called ``\textit{subreddits}.'' 
For example, multiple subreddits focus on Decentraland, namely \texttt{r/decentraland} for
general discussion, and \texttt{r/MANACOIN} for the MANA token.
Figure~\ref{fig:reddit_diagram} illustrates how users interact on Reddit.
Discussions start on Reddit with a user 
submitting a post related to the subreddit's theme. These posts, or ``\textit{submissions}'' following standard 
terminology, include a title and a 
body of text.
Other users can subsequently continue the discussion by responding to the submission. These responses only consist of a body of text (the title is that of the original submission) and are called ``\textit{comments}.''
An important feature here is that administrators and content moderators for subreddits have the authority to remove posts (submissions and comments) that violate terms of use and/or rules for subreddits.
Removal rules may vary across subreddits, but generally, inappropriate 
content such as scams, harassment, and more generally, posts considered 
abnormally rude are taken down from the platform.

We collect submissions and comments from 2019 through 2022 using the Pushshift Reddit
dataset~\cite{Baumgartner_Zannettou_Keegan_Squire_Blackburn_2020,
Proferes_Jones_Gilbert_Fiesler_Zimmer_2021}.
In this dataset, removed posts are flagged as ``removed'' -- while 
the text content is deleted, we still have access to metadata.

\subsection{Descriptive statistics}
\paragraph{Decentraland}
\begin{table}[htbp]
  \centering
    \caption{ \label{tab:basic_stats} Descriptive statistics for LAND and EST tokens.}
    \begin{adjustbox}{width=\textwidth,center}
    \begin{threeparttable}
      \begin{tabular}{llcccclcccclcccclcccclccc} \toprule \toprule
         & &
         \multicolumn{4}{c}{2019} & & 
         \multicolumn{4}{c}{2020} & & 
         \multicolumn{4}{c}{2021}
         \\
         \cmidrule{3-6} \cmidrule{8-11} \cmidrule{13-16}
         & &
         Q1 & Q2 & Q3 & Q4 & & 
         Q1 & Q2 & Q3 & Q4 & & 
         Q1 & Q2 & Q3 & Q4 \\ 
         \midrule 
         \multicolumn{2}{l}{\textbf{Number of transactions (LAND)}} \\
         & Listing &
         7,616 & 6,522 & 2,813 & 2,928 & &
         7,629 & 3,386 & 2,323 & 2,241 & &
         3,361 & 3,211 & 1,695 & 4,164
         \\
         & Sales (Decentraland) &
         973 & 500 & 364 & 315 & &
         819 & 257 & 424 & 380 & &
         952 & 684 & 547 & 1,553
         \\
         & Sales (Third-party marketplace) &
         20 & 7 & 1 & 5 & &
         23 & 69 & 53 & 18 & &
         135 & 403 & 450 & 1,537
         \\
         & Bids placed &
         2,632 & 2,897 & 1,630 & 2,238 & &
         3,517 & 1,311 & 392 & 499 & &
         365 & 444 & 230 & 603
         \\
         & Bids accepted &
         96 & 195 & 93 & 99 & &
         330 & 173 & 53 & 56 & &
         51 & 65 & 8 & 52
         \\
         \multicolumn{2}{l}{\textbf{Number of transactions (EST)}} \\
         & Listing & 
         3,594 & 3,002 & 1,379 & 1,454 & &
         3,248 & 1,608 & 1,120 & 921 & &
         1,161 & 1,189 & 755 & 1,504
         \\
         & Sales (Decentraland) &
         236 & 120 & 107 & 69 & &
         161 & 89 & 51 & 94 & &
         177 & 124 & 124 & 191
         \\
         & Sales (Third-party marketplace) &
         1 & 0 & 0 & 1 & &
         2 & 0 & 0 & 1 & &
         1 & 1 & 0 & 4
         \\
         & Bids placed & 
         887 & 1,510 & 749 & 1,145 & &
         1,446 & 880 & 182 & 233 & &
         201 & 561 & 577 & 444
         \\
         & Bids accepted &
         23 & 86 & 48 & 36 & & 
         68 & 56 & 24 & 23 & &
         20 & 59 & 51 & 56
         \\
         \multicolumn{2}{l}{\textbf{LAND median price [$10^3$ USD]}} \\
         & Listing &
         1.2 & 1.4 & 0.9 & 0.8 & &
         1.2 & 0.9 & 1.6 & 1.5 & &
         6.5 & 9.0 & 6.6 & 19.8
         \\
         & Sales (Decentraland) &
         0.6 & 0.7 & 0.5 & 0.4 & &
         0.8 & 0.6 & 0.8 & 0.7 & &
         4.0 & 6.0 & 5.0 & 15.4
         \\
         & Sales (Third-party marketplace) &
         0.7 & 0.5 & 0.6 & 0.4 & &
         0.5 & 0.5 & 0.8 & 0.5 & &
         4.5 & 4.3 & 4.2 & 14.4
         \\
          & Bids placed &
          0.7 & 0.7 & 0.5 & 0.4 & &
          0.6 & 0.5 & 0.8 & 0.7 & &
          3.2 & 5.5 & 4.2 & 10.0
          \\
         \multicolumn{2}{l}{\textbf{Trading volume$^{*}$ [$10^6$ USD]}} \\
         & LAND -- Sales (Decentraland) &
         0.8 & 0.5 & 0.3 & 0.2 & &
         0.8 & 0.2 & 0.4 & 0.4 & &
         4.7 & 4.6 & 3.1 & 26.5
         \\
         & LAND -- Sales (Third-party marketplace) &
         0.0 & 0.0 & 0.0 & 0.0 & &
         0.0 & 0.1 & 0.1 & 0.0 & &
         0.7 & 3.2 & 3.8 & 24.2 
         \\
         & EST -- Sales (Decentraland) &
         0.9 & 0.6 & 0.3 & 0.2 & &
         0.6 & 0.4 & 0.2 & 0.3 & &
         4.1 & 4.5 & 3.2 & 17.8
         \\
         & EST -- Sales (Third-party marketplace) &
         0.0 & 0.0 & 0.0 & 0.0 & &
         0.0 & 0.0 & 0.0 & 0.1 & &
         0.1 & 0.0 & 0.0 & 0.0
         \\
         \midrule
         & &
         \multicolumn{4}{c}{2022} & &
         \multicolumn{4}{c}{2023} & & 
         \multicolumn{4}{c}{5-year Stats.}\\ 
         \cmidrule{3-6} \cmidrule{8-11}
         & &
         Q1 & Q2 & Q3 & Q4 & &
         Q1 & Q2 & Q3 & Q4 \\ \midrule 
         \multicolumn{2}{l}{\textbf{Number of transactions (LAND)}} \\
         & Listing &
         3,018 & 2,012 & 1,282 & 794 & &
         1,211 & 475 & 509 & 449 & & 
         \multicolumn{4}{c}{57,639}
         \\
         & Sales (Decentraland) &
         1,010 & 477 & 214 & 212 & &
         167 & 81 & 88 & 93 & &
         \multicolumn{4}{c}{10,110}
         \\
         & Sales (Third-party marketplace) &
         1,197 & 724 & 364 & 395 & &
         335 & 221 & 190 & 284 & & 
         \multicolumn{4}{c}{6,431}
         \\
         & Bids placed &
         458 & 559 & 402 & 345 & &
         125 & 100 & 75 & 44 & & 
         \multicolumn{4}{c}{18,866}\\
         & Bids accepted &
         38 & 133 & 20 & 13 & &
         19 & 22 & 8 & 2 & & 
         \multicolumn{4}{c}{1,526}
         \\
         \multicolumn{2}{l}{\textbf{Number of transactions (EST)}} \\
         & Listing & 
         1,053 & 1,047 & 933 & 511 & &
         771 & 288 & 374 & 322 & & 
         \multicolumn{4}{c}{26,234}
         \\
         & Sales (Decentraland) &
         155 & 109 & 74 & 67 & &
         35 & 23 & 37 & 35 & & 
         \multicolumn{4}{c}{2,078}
         \\
         & Sales (Third-party marketplace) &
         22 & 31 & 40 & 47 & &
         6 & 0 & 6 & 5 & & 
         \multicolumn{4}{c}{168}
         \\
         & Bids placed & 
         1,112 & 411 & 1,131 & 1,486 & &
         241 & 305 & 276 & 124 & & 
         \multicolumn{4}{c}{13,901}
         \\
         & Bids accepted &
         84 & 43 & 42 & 42 & &
         32 & 18 & 16 & 8 & & 
         \multicolumn{4}{c}{835}
         \\
         \multicolumn{2}{l}{\textbf{LAND median price [$10^3$ USD]}} \\
         & Listing &
         17.3 & 8.9 & 5.1 & 2.9 & &
         3.0 & 2.0 & 1.3 & 1.4 & & 
         \multicolumn{4}{c}{2.0}\\
         & Sales (Decentraland) &
         15.0 & 6.3 & 3.4 & 2.2 & &
         2.1 & 1.3 & 0.9 & 0.8 & & 
         \multicolumn{4}{c}{3.1}\\
         & Sales (Third-party marketplace) &
         13.3 & 4.7 & 2.8 & 2.1 & &
         1.8 & 1.0 & 0.7 & 0.7 & & 
         \multicolumn{4}{c}{5.4}
         \\
          & Bids placed & 
          12.9 & 2.9 & 2.7 & 2.1 & &
          2.0 & 1.6 & 0.7 & 0.7 & & 
          \multicolumn{4}{c}{0.7}
          \\
         \multicolumn{2}{l}{\textbf{Trading volume$^{*}$ [$10^6$ USD]}} \\
         & LAND -- Sales (Decentraland) &
         16.8 & 3.5 & 0.9 & 0.5 & &
         0.4 & 0.1 & 0.1 & 0.1 & &
         \multicolumn{4}{c}{64.9}\\
         & LAND -- Sales (Third-party marketplace) &
         21.1 & 4.4 & 1.2 & 16.2 & &
         6.3 & 0.3 & 0.2 & 0.2 & &
         \multicolumn{4}{c}{82.0}
         \\
         & EST -- Sales (Decentraland) &
         14.7 & 4.0 & 1.4 & 0.8 & &
         0.3 & 0.1 & 0.2 & 0.3 & & 
         \multicolumn{4}{c}{55.0}\\
         & EST -- Sales (Third-party marketplace) &
         0.0 & 0.1 & 0.0 & 0.0 & &
         0.0 & 0.0 & 0.0 & 0.0 & & 
         \multicolumn{4}{c}{0.5}
         \\
         \bottomrule \bottomrule
      \end{tabular}
      \begin{tablenotes}
        \item[*] We exclude Decentraland sales whose price is higher than $10^{10}$ MANA as outliers.
      \end{tablenotes}
    \end{threeparttable}
  \end{adjustbox}
\end{table}

Table~\ref{tab:basic_stats} summarizes descriptive statistics about our
parcel transaction dataset. From the number of transactions and trading
volume, the number of sales transactions for LAND tokens is 5--8 times
larger than that of EST tokens in late 2021, while their trading volumes
are in the same order. Intuitively, this makes sense: EST prices are
typically several times higher than LAND tokens since they hold multiple
parcels. Buyers with smaller budgets, presumably individuals, are thus
more likely to have purchased LAND tokens, while considerably fewer
investors have enough purchasing power to acquire EST tokens.

Second, in the Decentraland LAND marketplace, the fraction of seller listings
(row~1) that result in a sale (row~2) is higher than the
fraction of (auction) bids (row~4) that result in successful purchase (row~5),
in most of our measurement period. 
Intuitively again, desirable parcels are more likely to be sold at or 
close to seller asking price. 
On the other hand, if a parcel is only sold through the bidding process, it probably evidences 
a discrepancy between the seller's expectations, and 
the buyers' willingness-to-pay, making successful transactions rarer.

More generally, price distribution follows an expected order: listing prices are higher
than direct sale prices, which are themselves higher than bidding
prices. 
Among direct sales, those taking place on the Decentraland
platform register higher prices than those on third-party marketplaces.
This makes sense, as most desirable parcels generally immediately sell
on Decentraland, while less popular land plots have to be shopped around
more extensively, including on third-party websites. Direct sales
registering for a higher price than accepted buyer bids is also unsurprising, as
discussed above. Auctions typically are a ``best offer,'' whereas direct
sales are much closer to the asking price.

Finally, starting in 2021, land prices show very large swings. The
median price of a listing is in the order of USD~1,000 at the beginning
of 2021, and jumped to close to USD~20,000 at the end of the year;
actual sales prices went from USD~700 or so to about 15,000 USD. We
note in particular a three-fold increase in sales price in Q4 only.
However, the market quickly cooled down: by Q3 of 2022, sales prices
had gone back down to about USD~3,000, and are back below USD~1,000 \dkEraseTwo{at the time of writing}\dkAddTwo{in} 2023 Q4. In short, these prices movements seem to
indicate that LAND tokens experienced a speculative bubble, which we
will study further. 

\begin{figure}[htbp]
    \centering
    \includegraphics[width=0.9\columnwidth]{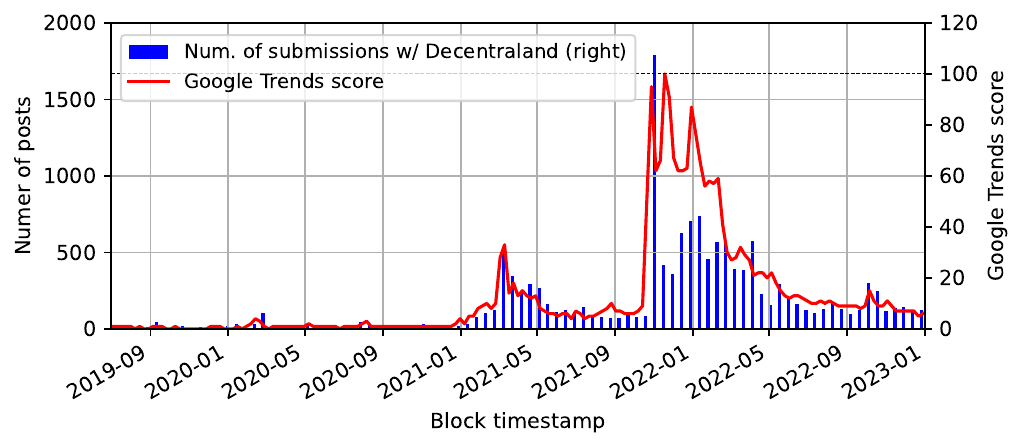}
    \caption{LAND trading volume and number of Reddit posts.
      \dkAddTwo{
        The blue bars and red line represent the number of submissions containing the word "Decentraland" over two weeks and the Google Trend score for one week, respectively.
    }}
    \label{fig:num_posts}
\end{figure}

\paragraph{Reddit}
Figure~\ref{fig:num_posts} shows the number of submissions on Reddit referring to ``Decentraland'' and the Google Trends score for the word ``Decentraland''  between Jan.~1, 2019 and Dec.~31, 2022.
The number of submissions is highly correlated with Google Trends score (Pearson correlation coefficient = 0.75).
This indicates that the Reddit dataset mirrors the general interest of web users in Decentraland.

The substantial increase in Oct. 2021 is due to a bot-like submission recommending investment in Decentraland-related assets in a specific marketplace (see Appendix~\ref{appx:reddit_exclusion}).
Therefore, we exclude these submissions from the analysis in subsequent sections.

\section{Methods}
\label{sec:dl_method}
To analyze the popularity and trading dynamics of Decentraland, we
first explore topics discussed on Reddit. We then leverage blockchain
records pertaining to Decentraland to carry out a regression analysis
and compute profits and losses.
Throughout this study, we use daily cryptocurrency
price data from Yahoo! Finance to convert land prices
denominated in cryptocurrencies to their corresponding USD
value.\footnote{\url{https://finance.yahoo.com/crypto/}.}

\paragraph{Topic-modeling analysis of Reddit posts}
We first investigate why Decentraland parcels gained popularity in 2021 from a topic-modeling analysis of Reddit posts. 
Specifically, we use the Gibbs-Sampling Dirichlet Multinomial Mixture 
(GSDMM) method \cite{Yin_Wang_2014}---a topic-modeling method frequently used for short text such as social media posts \cite{Qiang_Qian_Li_Yuan_Wu_2022}---to identify topics of submissions referring to ``Decentraland'' either in the title or in the body.
(See Appendix~\ref{appx:reddit_exclusion} for the details of GSDMM analysis.)
For each topic identified by GSDMM, we count the number of submissions that
match that topic. This allows us to identify what users are most interested
in when discussing Decentraland, in particular, when the prices of 
Decentraland-related items (e.g., parcels) increases. 

Starting in the second half of 2022, when prices start to fall 
significantly, an increasing fraction of comments are removed. This 
may give us clues about user frustration (as angrily venting may result 
in posts violating 
terms of service). Specifically, topic modeling allows us
to better understand 
what users are most vociferous about. 

\paragraph{Land price analysis}
We next discuss our regression analysis on parcel prices. 
Our model is based on an economic theory that explains real-estate valuations, called \textit{bid-rent theory} \cite{Alonso_1960, Alonso_1964, Fujita_Thisse_2002}.
The fundamental idea is that the closer to a central business district (CBD), the higher the land price.
This is because retail businesses view proximity to the CBD as more profitable than remote areas (due to the better access to customers), which in turn leads to increased competition between possible tenants and results in higher real estate valuation.

Bid-rent theory assumes the existence of a single center, which is the case in Decentraland. 
Users spawn from the center of the map (i.e., the Genesis Plaza in the center of Figure~\ref{fig:decentraland_map}). So,
businesses close to the Genesis Plaza should theoretically have more
exposure to customers. A notable wrinkle compared to the physical world
is that users can ``teleport'' to any parcel on the map once they spawn.
In practice, though, we hypothesize that most users are primarily
exploring the world by wandering around, mimicking the behavior of their
real-world counterparts.
On the other hand, parcel owners can develop their parcels very much 
like they would in the real-world: they can construct buildings, or host 
businesses such as NFT galleries \cite{Decentraland_2021_gallery} or 
casinos. 
\dkAdd{
In fact, our analysis of the number of visitors to parcels indicates that the hypothesis holds true for Decentralnad.
See Appendix~\ref{appx:reg_suppl} for the details.
}
Therefore, distance could still be an important metric in determining the land price both for sellers and buyers.
(We will prove this is indeed the case in Section~\ref{sec:decentraland_result}.)

We use this assumption to build an empirical regression, following Gupta et al.~\cite{Gupta_Mittal_Peeters_Van_Nieuwerburgh_2022}:
\begin{align}
\label{eq:reg_formula}
\log P_{i,n} = & c + \sum_t \left[\alpha_t + \delta_t \log \left(1 + D_i \right) \right] Quarter_{t,n} \\
& + \sum_{J}\beta_J I_{J,i} + \epsilon_n \ , \ \   \epsilon_n \sim \mathcal{N}(0, \sigma^2) \, , \nonumber
\end{align}
where $P_{i,n}$ is the price of parcel $i$ at the $n$-th sales or listing,
$\alpha_t$, the overall shift
in land price at $t$-th quarter ($t = \mbox{2019-Q1}, \dots,
\mbox{2022-Q4}$), $D_i$, the 
Euclidean distance of parcel $i$ from the
center of Decentraland map, and $Quarter_{t, n}$ an indicator variable
denoting whether the $n$-th sale/listing happened in the $t$-th quarter (i.e.,
$Quarter_{t, n} = 1$ if the n-th sales/listing happened in $t$-th quarter,
$0$ otherwise). $I_{J, i}$ is the vector of $J$ control variables of
interest for parcel $i$, which we will introduce in the next paragraph.
Last, the regression variable $\delta_t$ denotes the spatial dependence of
land prices. In particular, a significantly negative $\delta_t$ means
that parcels close to the center of the Decentraland map will be more
expensive than others.

For our regression analysis, we will limit ourselves to listing and
sales (i.e., purchase) prices for the LAND tokens on the Decentraland marketplace,
and will not include third-party markets and EST token transactions in this analysis.
We motivate this choice with four reasons: 1) sales on Decentraland dominate
other categories of transactions (see Table~\ref{tab:basic_stats});
2) transactions on Decentraland have no ambiguity in price breakdown,
compared to those in third-party marketplaces that may include a LAND
token as part of bulk offerings, possibly containing other, unrelated
NFTs (see Sec.~\ref{sec:dl_dataset} for the details); 3) prior to 2021, third-party marketplace sales of LAND token
were rare; 4) a number of bids appear to be
unrealistic: they are several orders of magnitude lower than the median
asking price, have therefore no chance to be accepted, and would only
create noise in the regression.

\paragraph{Control variables}
We have four control variables of interest. 
First, closeness to a street can be advantageous to bring in customer traffic to the parcel, thereby commandeering a higher land price.
Moreover, in Decentraland, the profile page of a parcel prominently highlights information about its closeness to streets, districts, and Genesis Plaza, which may impact the parcel's price.
We capture these effects with three control variables: 
\begin{equation} 
  I_{Pub,i} = \left\{
  \begin{array}{ll}
    1 & \mbox{if the closeness to $Pub$}\\
      & \mbox{is displayed on}\\
      & \mbox{the profile page of $i$-th parcel,}\\
    0 & \mbox{otherwise.}
  \end{array}
  \right .
\end{equation} 
where $Pub$ denotes one of \{road, district, Plaza\}.

We also notice that a limited number of parcels very close to the center of the map are exceptionally expensive compared to others and show a discrepancy in prices (see Appendix~\ref{appx:reg_suppl}).
We account for this with a fourth control variable $I_{Close, i}$:
\begin{equation}
  I_{Close, i} = \left\{
    \begin{array}{ll}
      1 & \mbox{if $D_i < 20$,}\\
      0 & \mbox{otherwise.} 
    \end{array}
    \right .
\end{equation}

$I_{Close}=1$ for only 518 out of 44,480 ownable parcels (1.16\%), so, for most parcels, any adjustment due to $I_{Close}$ will not be significant. 

\paragraph{Profit and loss for land trading}
Finally, we calculate the profit and loss for each unique Ethereum
account involved in parcel trading. To do so, we will consider all LAND token sales
transactions, including bids accepted in the Decentraland marketplace
and sales in third-party marketplaces to track transfer records as completely as possible.
We only exclude accounts that have transferred or received LAND tokens for free (to avoid reidentifying owners). 
For bulk sales, we estimate the price per parcel to simply be the total price divided by the number of NFT items involved.
The inclusion of EST token transactions does not change results while a little noise is introduced.

\section{Analysis of Reddit posts}
\label{sec:reddit_analysis}
\begin{figure}[htbp]
    \centering
    \begin{subfigure}[b]{0.3\textwidth}
    \centering
        \includegraphics[width=0.95\textwidth]{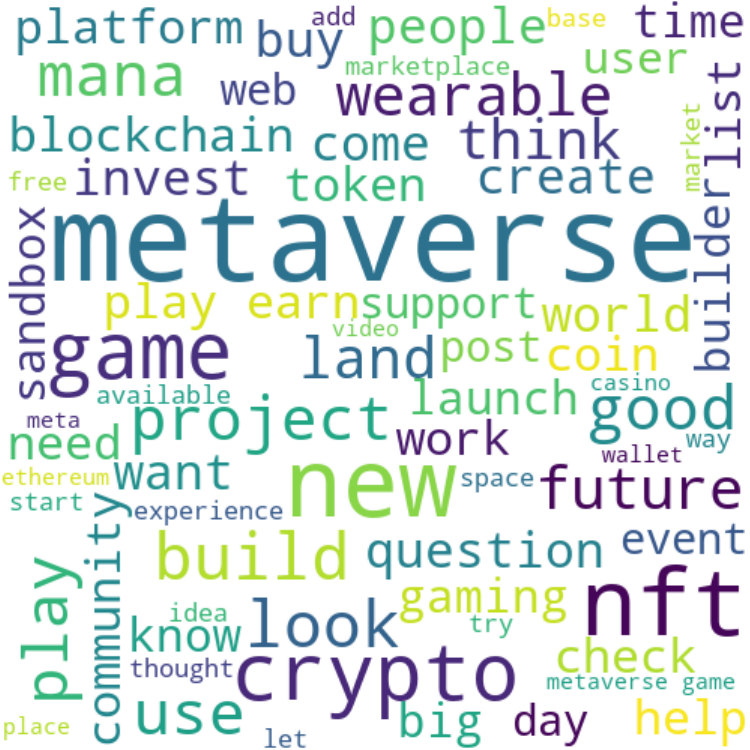}
        \caption{\label{subfig:primary_topic_DMM} First topic}
    \end{subfigure}
    \begin{subfigure}[b]{0.3\textwidth}
    \centering
        \includegraphics[width=0.95\textwidth]{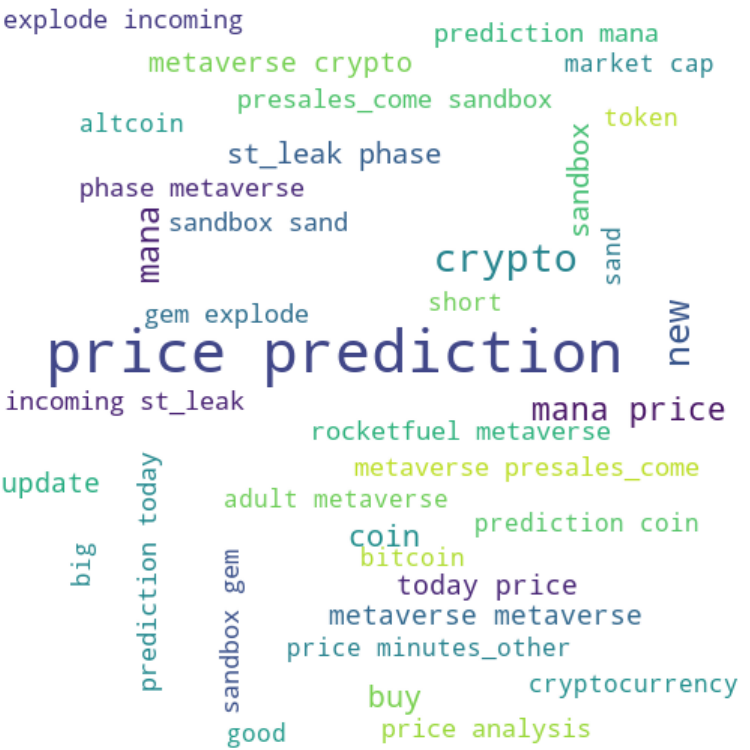}
        \caption{\label{subfig:secondary_topic_DMM} Second topic}
    \end{subfigure}
    \begin{subfigure}[b]{0.3\textwidth}
    \centering
        \includegraphics[width=0.95\textwidth]{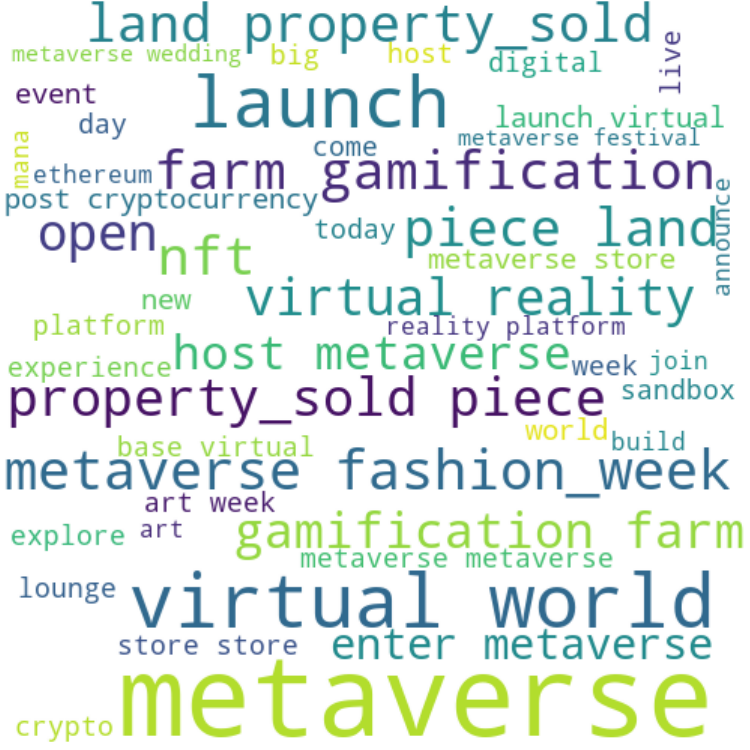}
        \caption{\label{subfig:third_topic_DMM} Third topic}
    \end{subfigure}
    \caption{ \label{fig:topics_DMM} The word cloud of the three most salient topics according to GSDMM.
    The size of words reflects their relative importance within the topic.}
\end{figure}

Figure~\ref{fig:topics_DMM} shows word clouds for the three most 
common topics broached when discussing Decentraland -- as categorized 
by the Gibbs-Sampling Dirichlet Multinomial Mixture (GSDMM) method.
The most prominent topic is about 
the metaverse (not specific to Decentraland), NFTs, and cryptocurrencies; the second most prominent topic is investment advice regarding assets associated with metaverse projects; and the third most prominent topic concerns in-world events and features of Decentraland, such as fashion week,\footnote{\scriptsize \url{https://decentraland.org/blog/announcements/tradition-and-innovation-collide-decentraland-metaverse-fashion-week-2023}} art week,\footnote{\scriptsize \url{https://decentraland.org/blog/announcements/metaverse-art-week-2023-the-metaverse-is-dead-long-live-the-metaverse}} and stores.
These topics comprise 26.6\%, 15.5\%, and 13.6\% of submissions, respectively.
In short, Reddit users discussed first and foremost 
Decentraland in terms of a metaverse project leveraging blockchain technology, 
but also showed a secondary 
interest in profiting from the investment in metaverse-related items, and, to a lesser extent, an interest in features specific 
to Decentraland.

\begin{figure}[htbp]
    \centering
    \includegraphics[width=0.95\textwidth]{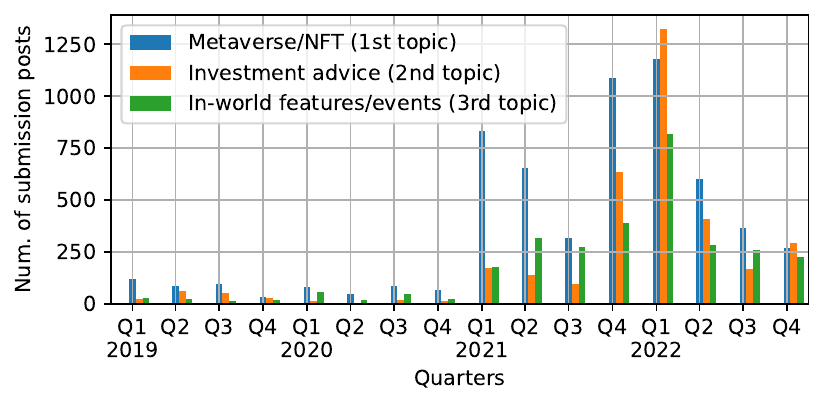}
    \caption{\label{fig:num_submission_DMM}Number of submission posts for the three most popular Decentraland topics under DMM analysis.}
\end{figure}

Figure~\ref{fig:num_submission_DMM} shows the temporal distribution of submissions categorized into the aforementioned topics.
We observe interesting differences. 
The most salient topic, i.e., metaverse and NFT, increased in early 2021 and 
again in the 4th quarter of 2021, which were the times when NFTs gained 
substantial popularity, and Facebook rebranded itself as ``Meta'' in the 
hope that it would be the metaverse industry leader.
In particular, the 4th quarter of 2021 features ten times more 
submissions than a year prior,
indicating that Decentraland abruptly gained substantial popularity on 
Reddit, presumably due to the rising hype about NFTs and the metaverse.
The second most popular topic (investment advice) shows an even more 
interesting distribution.
In early 2021, the number of submissions categorized under 
this particular topic was the smallest among the main three topics of
discussion. 
However, the number of investment-related posts experienced a significant
increase and became the most actively discussed topic in the first quarter 
of 2022, denoting increasing excitement about investment opportunities.
This coincides with the time when the metaverse gained renewed attention 
after Facebook's announcement. A possible explanation is that that 
event sparked conversations about profiting from metaverse-, and by extension, Decentraland-related assets.
In fact, the trading volume of Decentraland parcels is highly correlated 
with the number of submissions featuring these 
two most popular topics; the Pearson correlation coefficients are 
0.74 and 0.83, respectively.
On the other hand, submissions related to the third topic---in-world 
features---increased more moderately. 
Figure~\ref{fig:num_submission_DMM} shows that Decentraland features 
were more actively discussed than Decentraland-related investments before 
the end of 2021 -- but the tide turned in late 2021, precisely as 
hype surrounding the metaverse started to build up. This suggests that 
this renewed interest in the metaverse primarily translated into investment-related  (rather than feature-related) discussions, casting doubts 
user interest in actively participating in and building the platform. 

\begin{figure}[htbp]
    \centering
    \begin{subfigure}[b]{0.475\textwidth}
    \centering
        \includegraphics[width=\textwidth]{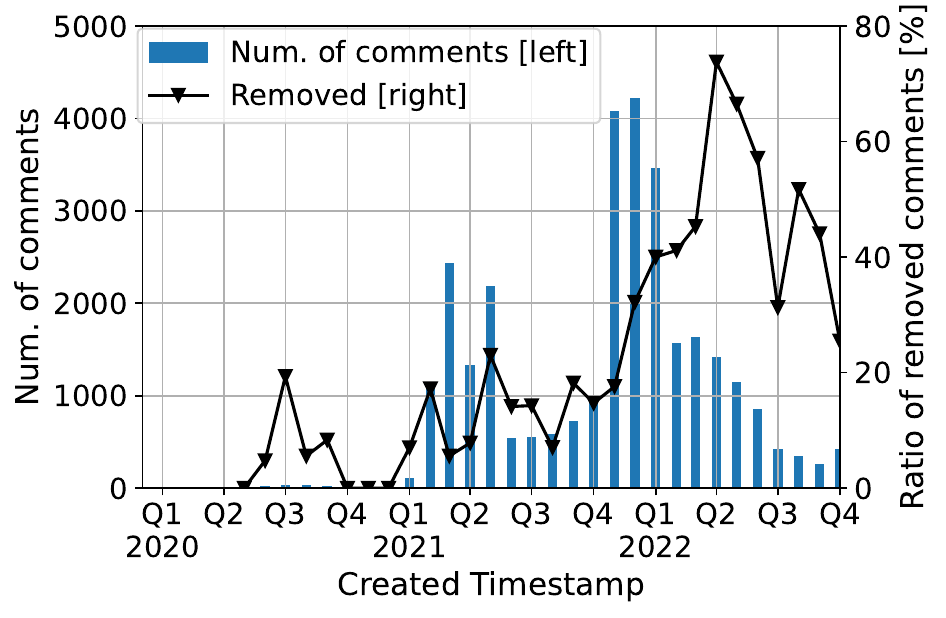}
        \caption{\label{subfig:primary_toxicity} Metaverse/NFT (most popular topic)}
    \end{subfigure}
    \begin{subfigure}[b]{0.475\textwidth}
    \centering
        \includegraphics[width=\textwidth]{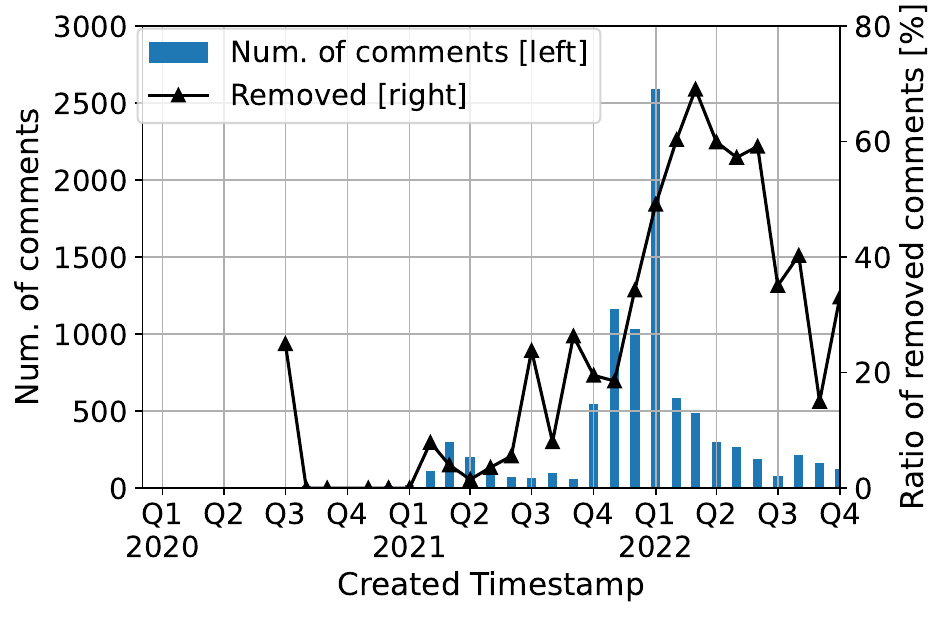}
        \caption{\label{subfig:secondary_toxicity} Investment advice (second most popular topic)}
    \end{subfigure}
    \caption{ \label{fig:toxicity} Comments on submissions categorized into the two main topics of discussion, and ratio of removed comments over time.}
\end{figure}

Figure~\ref{fig:toxicity} shows the number of top-level comments (i.e., replies to submissions) over time for the two most popular topics of discussion.
As expected, the evolution over time of the number of comments tracks closely that of submissions 
(Figure~\ref{fig:num_submission_DMM}). More interestingly, the ratio of 
removed comments increases sharply starting in Q1 2022 -- both for 
Metaverse-related comments and for investment advice -- reaching more than 
60\% in the first half of 2022.
For comparison, the average removal rate on all of Reddit is 21.8\% \cite{Jhaver_Bruckman_Gilbert_2019}. 
This surge in removed comments is temporally correlated with the
number of
transactions and prices of Decentraland land parcels plummeting 
(Table~\ref{tab:basic_stats}). While we do not have access to the text 
body of these removed comments, it seems highly likely that the sheer 
number of removed comments likely evidences user 
frustration about the poor performance of Decentraland assets.

\section{Analysis of Decentraland data}
\label{sec:decentraland_result}
We next conduct the regression analysis of LAND token prices and then
explore the profits and losses incurred by market participants.

\subsection{Regression analysis}
\label{subsec:regression_result}
Figure~\ref{fig:spatial_dep} shows the spatial dependence of land prices ($\delta$) (see Appendix~\ref{appx:reg_suppl} for details).\footnote{We excluded a small fraction of observations whose studentized residuals are more than 3.00 as outliers \cite{Weisberg_2005}.} The grayed area corresponds to a 18-month window centered on 2021 Q4, which is when the hype bubble we identified earlier peaked.  We first observe a clear 
difference between the ``pre-bubble'' situation from 2019 to early 2021 and the 2021--2022 bubble period. Specifically, before 2021, 
all prices show significant negative dependence -- $\delta_t$ is around $-0.4$ or even lower for listing prices, indicating that sellers definitely account for location in their asking prices; buyers do, too, to a slightly lower, but nevertheless significant extent ($\delta_t \in [-0.2, -0.4]$).
Concretely, parcels at a distance of 20 units from the Genesis Plaza ($\log D_i \simeq 3$) sell roughly for twice as much as the parcels at the edge of the map ($\log D_i \simeq 5)$.

However, Figure~\ref{fig:spatial_dep} shows a steep decline in the influence of geographical distance on purchase prices starting in 2021 Q2, when land prices per parcel increased to 5,000 USD, and NFT started to gain substantial popularity.
$\delta_t$ even becomes statistically insignificant at the 5\% level in 2021 Q4, meaning geographical distance has essentially no bearing on sales price at that time; this is when Decentraland abruptly gained increased attention as a metaverse platform.
Interestingly, the trend somewhat reverses after the hype bubble
has subsided (i.e., $\delta_t$ becomes statistically significant
again, albeit less influential than pre-bubble). 2023 Q3 is an
exception ($\delta_t \approx 0$), but this is probably due to
the fact that the platform has basically become inactive: from
Table~\ref{tab:basic_stats}, there were only 88 sales in that quarter,
compared to several hundreds or more in earlier periods.

In an effort to check whether these findings generalize to other 
platforms, we analyzed the sales price of land parcels in The Sandbox, 
another major metaverse project leveraging blockchain technology.
We found a similar decrease in the influence on prices of the distance from the
center (see Appendix~\ref{appx:reg_suppl} for the details).

\begin{figure}[htbp]
    \centering
    \includegraphics[width=0.95\columnwidth]{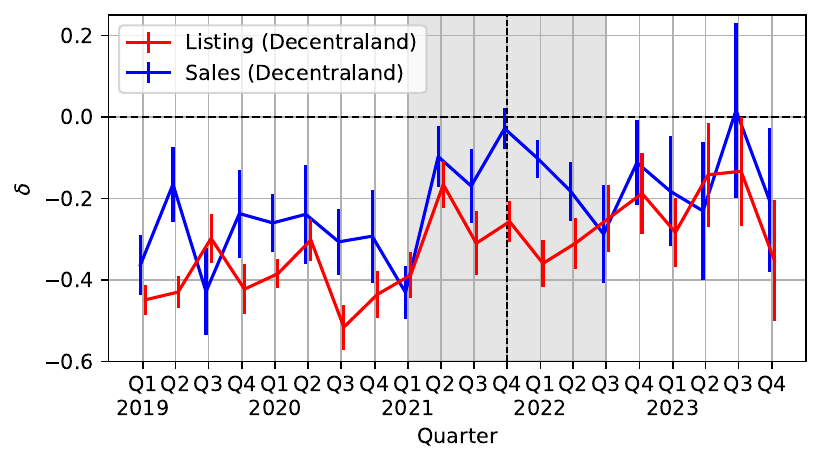}
    \caption{The spatial dependence of land prices ($\delta$) for quarters in 2019-2023.
    The error bars represent 95\% confidence intervals (CIs).
    The grey-colored area shows the 18-month interval surrounding the ``peak hype'' of 2021 Q4.
    }
    \label{fig:spatial_dep}
\end{figure}

\subsection{Analysis of parcels' value for commercial purpose}
\label{subsec:commercial_purpose}
This section provides a deeper analysis of the inflating prices of Decentraland parcels in the bubble period.
Decentraland parcels may have two distinct values. 
First, as discussed above, parcels have value as investment assets: the increasing prices could provide parcel owners with an opportunity for short-term capital gain. 
The other value lies in the potential for commercial activity in Decentraland.
The surging popularity of Decentralnd would increase the number of active users, which is a significant chance for parcel owners to profit from commercial activities in Decentraland.
We consider the contribution of the value for commercial activities with the building construction history on Decentraland parcels.

\begin{figure}[htbp]
    \centering
    \includegraphics[width=\textwidth]{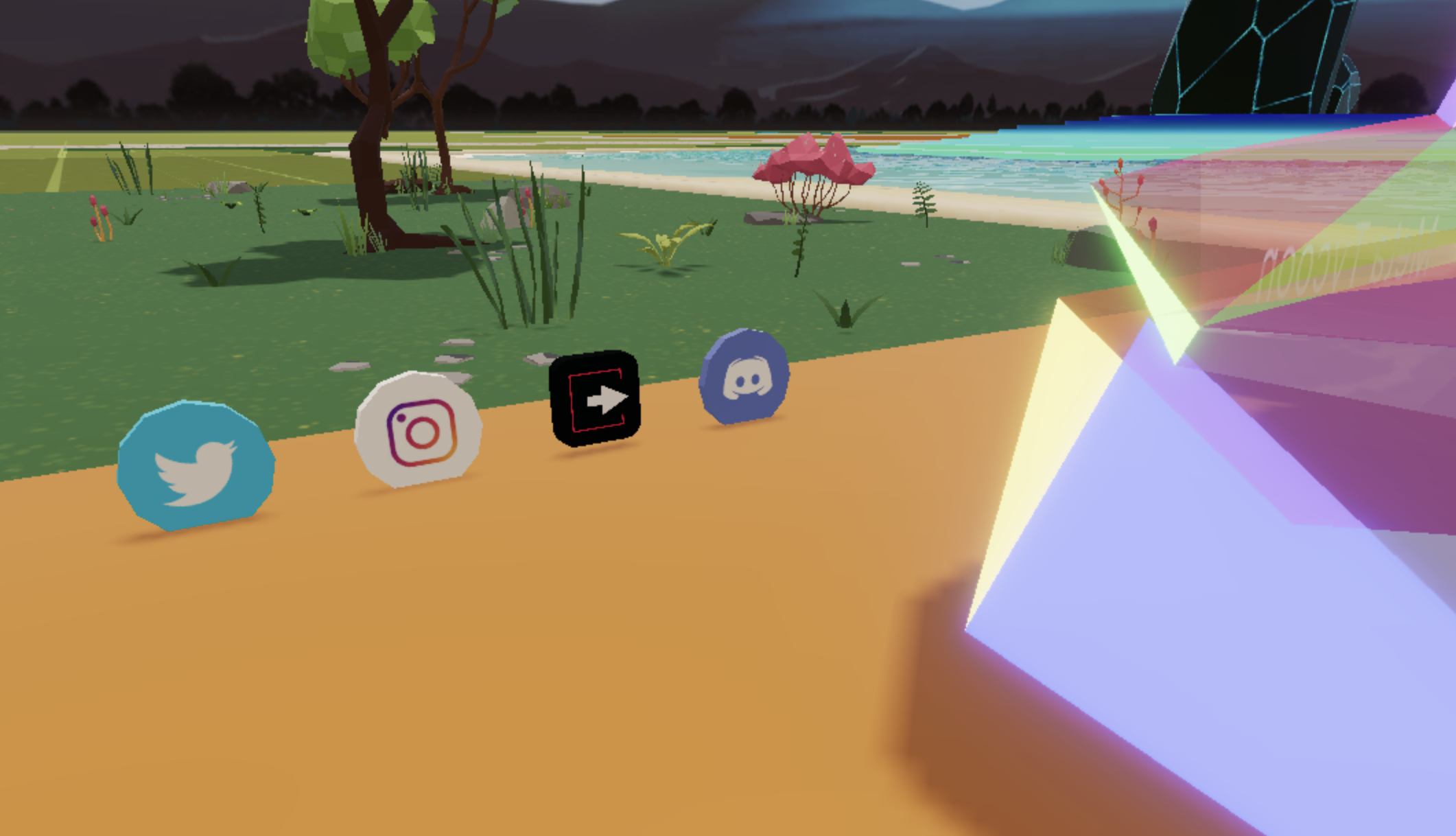}
    \caption{Snapshot of a Decentraland parcel with social media icons that have links to owner's social media handles advertising their created items.}
    \label{fig:decentra_advertise}
\end{figure}

The primary commercial use of Decentraland parcels is advertising creative products, particularly artworks published as NFT items. 
Parcel owners place social media icons with links to their handles, where they can promote their creations in more detail.
For example, Figure~\ref{fig:decentra_advertise} shows a snapshot of a Decentraland parcel that has the links to its owner's social media handles.
When a player visits the parcel, they can hover over the icons and click on the links to visit the owner's social media handles. We use the number of parcels that have links to social media profiles as an indicator of commercially used parcels and consider it as a measure of the prominence of commercial use of Decentraland parcels.

\begin{figure}[htbp]
    \centering
    \begin{subfigure}[b]{0.75\textwidth}
        \includegraphics[width=\textwidth]{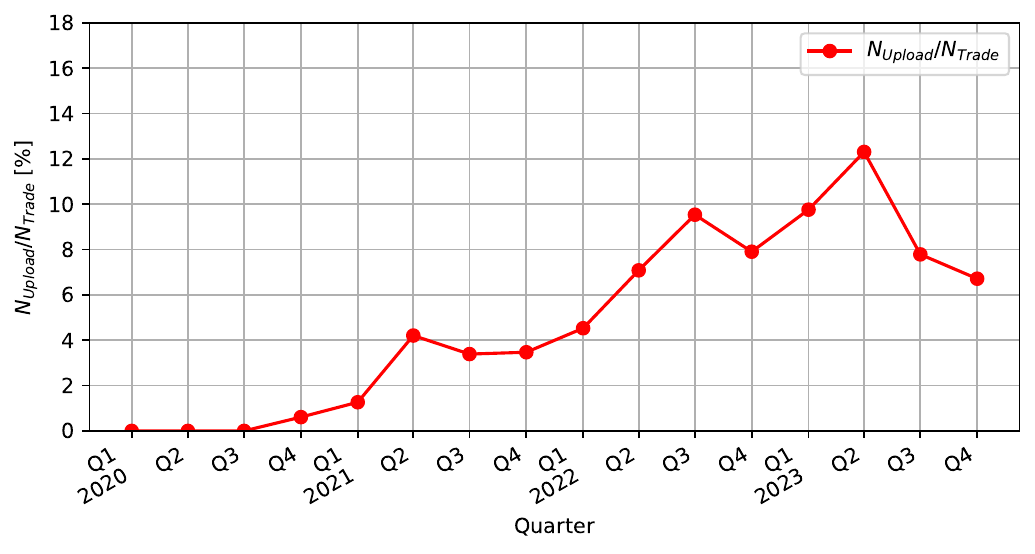}
        \caption{The number of uploaded scenes ($N_{upload}$) normalized by the number of parcel purchases ($N_{Trade}$).}
         \label{subfig:uploaded_scenes}
    \end{subfigure}
    \hfill
    \begin{subfigure}[b]{0.75\textwidth}
        \includegraphics[width=\textwidth]{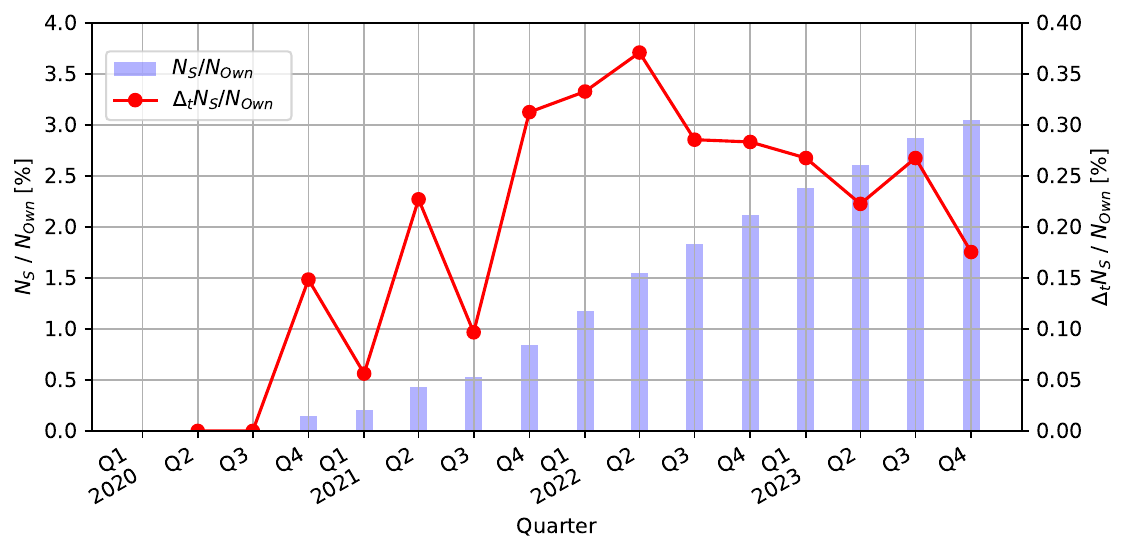}
        \caption{The number of developed parcels that have social media links normalized by the number of personally owned ones. The blue bars show the number of developed parcels that have social media links ($N_{s}$) normalized by the number of personally owned ones ($N_{own}$). The red line shows the increase per quarter ($\Delta_t N_s/N_{own}$).}
         \label{subfig:developed_parcels}
    \end{subfigure}
    \caption{\label{fig:development_with_social_media} Building statistics on Decentraland parcels.
    Figure~\ref{subfig:uploaded_scenes} shows the number of uploaded scenes that have social media links normalized by the number of parcel purchases.
    Figure~\ref{subfig:developed_parcels} shows the number of developed parcels that have social media links normalized by the number of personally owned parcels.
    }
\end{figure}

Figure~\ref{fig:development_with_social_media} illustrates the number of uploaded scenes (i.e., the appearance of a parcel) that have social media links normalized by the number of parcel purchases and the number of developed parcels that have social media links normalized by personally owned parcels.\footnote{The parcels do not belong to districts, Genesis Plaza, or roads.}
The normalization by parcel purchases helps us understand the proportion of parcel owners interested in promoting their creations compared to all owners.\footnote{The number of purchases represents the minimum number of times parcels have changed ownership.}
As for the normalized number of developed parcels (lower figure), we assume that developed parcels are never cleared to vacant lots once a scene is uploaded, even if their owners change. 
Although this is a bold assumption, it aligns with rational behavior: while the owner may alter the scene, there is no rational incentive to clear the land and leave it vacant, as doing so does not increase the parcel's value. 
This assumption implies that the number of developed parcels never decreases.

The figure depicts an interesting picture of commercial usage.
To begin, the upper figure illustrates that only a small percentage of owners (5--10\%) developed their holdings for advertisement. 
In particular, the ratio is only 4\% in the fourth quarter of 2021.
This suggests that they were not very interested in developing their properties to profit from commercial activities.
The lower figure further emphasizes this point by showing that only less than 1.0\% of properties were linked to their owners' social media profiles in the fourth quarter of 2021. 
This indicates that commercial utilization of properties is not the primary driver of the price surge in late 2021.

Another important point illustrated in the figure is that the increase in the ratio does not align with the inflating prices of Decentraland parcels (Table~\ref{tab:basic_stats}). 
While parcel prices peaked in the fourth quarter of 2021 and have continuously decreased since then, the ratios shown in Figure~\ref{fig:development_with_social_media} continue to increase until the second or third quarter of 2022. 
This indicates that the value of parcels for commercial use does not account for the change in parcel prices.
Instead, the shift in the number of parcels linked to social media accounts suggests a change in owners' motivation for purchasing parcels, from speculation in parcels to commercial purposes after the NFT and Metaverse frenzy subsides.

\paragraph{Ruling out alternative hypotheses} 
We explored various alternative explanations for this result, beyond a lack 
of enthusiasm toward commercial opportunities in virtual worlds.

First, parcels could be valued for the social activities they could host, 
rather than their commercial potential---if so, 
customer exposure would become 
a much less important factor, and geographic dependencies would become 
marginal.
However, the high parcel 
price during the booming period makes this alternative hypothesis 
highly unlikely: it is economically 
unreasonable that buyers would pay more than 15,000 USD for purely social 
activities (See Table~\ref{tab:basic_stats} for median sales prices).

Another possible explanation is that the CBD of Decentraland expanded to the point of saturation, making the distance from the center less important.
\dkAddTwo{Namely, if all parcels were developed from nearby to distant locations, thereby making remote areas attractive to users just like areas near the center, the distance from the center would be less crucial.}
Our analysis rules out the possibility.
We find that only 20-30\% of parcels were developed by the end of the fourth quarter of 2021\dkEraseTwo{, and the developments are concentrated in the central area} (see Appendix~\ref{appx:reg_suppl} for the details).
These data clearly reject the saturation hypothesis.

\begin{figure}[htb]
    \centering
    \includegraphics[width=0.95\columnwidth]{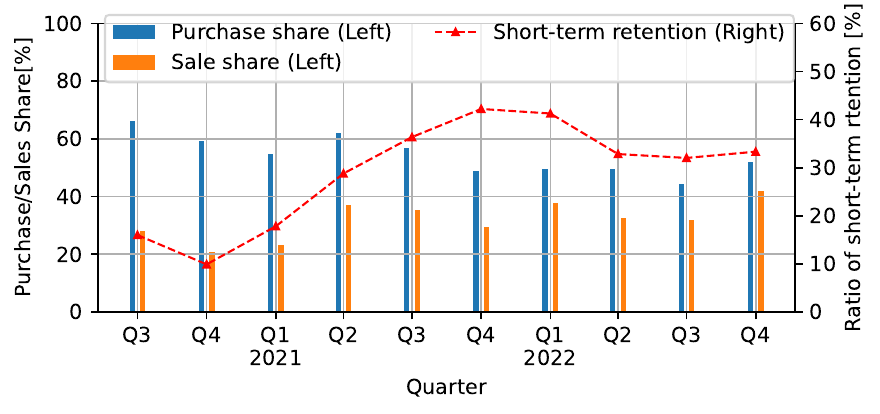}
    \caption{Sales and purchases of multiple-parcel buyers, and ratio of short-term ($<$ 2 weeks) holdings over parcels these buyers acquired in each quarter.}
    \label{fig:share_turnover}
\end{figure}

A third alternative interpretation is that 
many potential buyers were interested in commercial activity but were priced out of the market due to inflated parcel prices:
 the average
parcel listing price tripled from 6,000 USD to 20,000 USD in a quarter
(Table~\ref{tab:basic_stats}). 
However, 
there were still many wealthy enough buyers in the market.
Specifically, Figure~\ref{fig:share_turnover} shows three quantities related to buyers
of multiple parcels. 
The bar plots show the fraction of transactions in
which accounts that bought multiple parcels 
engaged (purchases in blue, sales
in orange). We do not observe any meaningful decrease in purchases
-- 48.9\% of sales in 2021 Q4 were to people buying more than one
parcel. 
This suggests that some buyers have a large enough budget to buy parcels close to the center: \dkAddTwo{the expected price of parcels close to the center is similar to that of remote parcels in 2021 Q4 as shown in Figure~\ref{fig:spatial_dep}.}
\dkEraseTwo{two parcels at a distance of 150 ($\log D_i\simeq 5$) are priced roughly as much as one parcel at a distance of 20 ($\log D_i \simeq 3$).}
Even more tellingly, the dashed line is the
percentage of parcels bought by buyers of multiple parcels in a quarter and resold within two weeks
or less (i.e., the ratio of short-term retention). We clearly observe
an increase in ``flipping'' activity during the bubble: more than 40\% of the parcels are sold within two weeks. 
In short, we can reject the hypothesis that \dkEraseTwo{interest} buyers are priced out 
of the market. Instead, a large number of those wealthier, influential investors do not appear to
invest in the long-term vision of a metaverse; instead, they
merely attempt to make quick profits, which explains why valuations
become disconnected from geographic considerations.

\begin{figure}[htb]
    \centering
    \includegraphics[width=0.95\columnwidth]{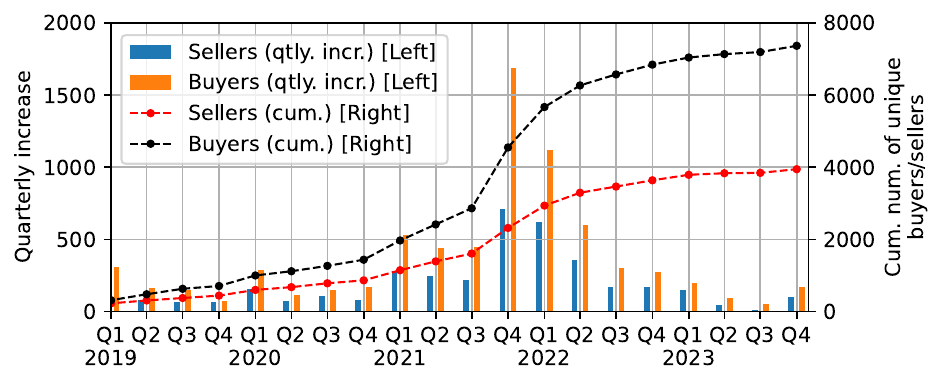}
    \caption{Cumulative number of unique sellers and buyers and its quarterly increase.}
    \label{fig:cum_account}
\end{figure}

Another significant change in market demographics is the large inflow of new entrants.
Figure~\ref{fig:cum_account} shows the cumulative number of unique sellers and buyers in quarters and its quarterly increase.
We see that there were 1,750 new buyers in 2021 Q4, \textit{three times} the number for the
previous quarter.
Also, new entrants to that market in that quarter account for more than one-third of the cumulative number of unique buyers.
Given our Reddit analysis, some of these buyers may be jumping on the metaverse bandwagon. We will see next whether this results in a profit.

\subsection{Land sale profitability}
\begin{figure}[htbp]
    \centering
    \includegraphics[width=0.95\columnwidth]{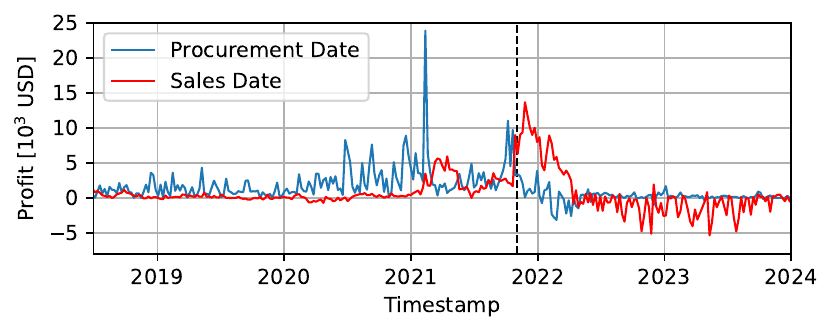}
    \caption{Profit landowners made based on when a parcel was bought (``procured'') or sold.
    The vertical black dashed line is the last week of Oct.~2021, when LAND price surged.
    }
    \label{fig:profit_procure}
\end{figure}

Figure~\ref{fig:profit_procure} shows landowner profit based on when
LANDs token were purchased and sold. For instance, an
account which buys LAND(s) in 2021 Q1 and sells it in 2021 Q2 is listed 
with a procurement date of 2021 Q1, and a sales date of 2021 Q2.

The figure shows a clear contrast between early adopters and the new
entrants who joined during the bubble. Looking at profit as a function
of the sales date, we observe that the largest profits ($\approx$ 10,000
USD) were for parcels sold in 2021 Q4 and 2022 Q1. After that, most
sellers actually incurred losses in the order of 1,000 USD on average.
When plotting profits as a function of the purchase date, we can see
that early adopters who bought parcels at the beginning of 2021, followed by purchasers in 2020 and just before the price increase in late 2021, reaped the highest profits. Conversely, those who joined during the bubble (and especially right after the peak of October 2021)
did not make significant profits---in fact, they barely broke even or
even lost money.

\begin{figure}[htbp]
    \centering
    \begin{subfigure}[t]{0.9\columnwidth}
        \includegraphics[width=\textwidth]{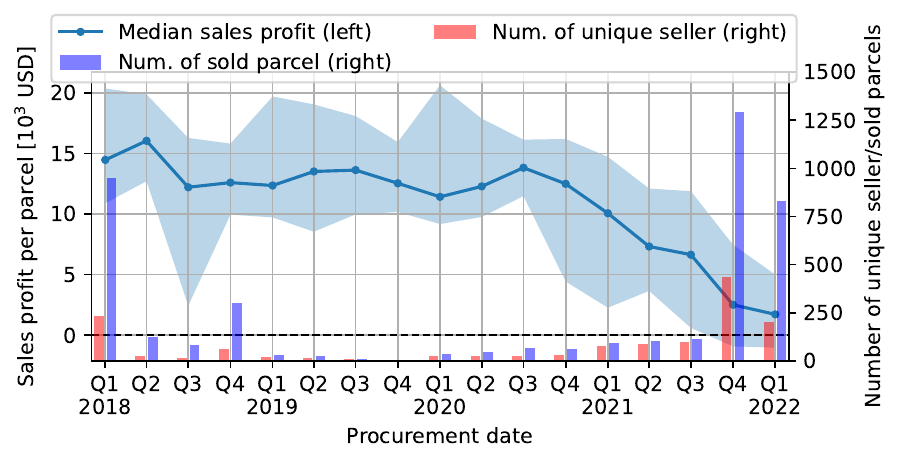}
    \end{subfigure}
    \begin{subfigure}[t]{0.9\columnwidth}
        \includegraphics[width=\textwidth]{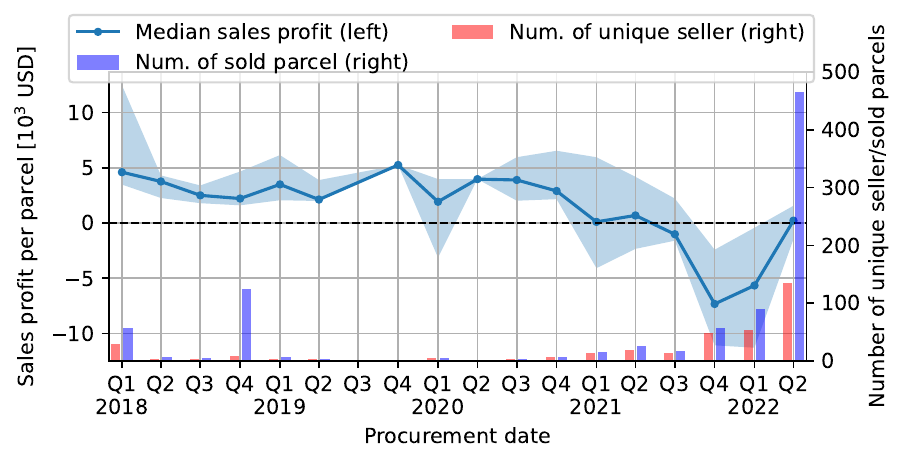}
    \end{subfigure}
    \caption{Median sales profit per parcel, number of unique sellers, and number of parcels sold during selected quarters for the period of 2021 Q4 -- 2022 Q1 (upper panel) and 2022 Q2 (lower panel).
    The light-blue area is the 15th-85th percentile.}
    \label{fig:ratio_buyer_seller}
\end{figure}

Figure~\ref{fig:ratio_buyer_seller} further substantiates this finding. It shows the median profit per parcel and the number of unique sellers for parcels sold during 2021 Q4 and 2022 Q1 (upper) and in 2022 Q2 (lower).
Early adopters made significant profits during the bubble
(upper panel); but for those who waited too long
(lower panel) did not make as much. More importantly, people
who joined after 2021 (i.e., new entrants possibly attracted by the
metaverse hype bubble) and sold in 2022 Q2 lost money. The number of unique
sellers (and sold parcels) also tells us that early adopters provided
somewhat constant liquidity before and during the bubble; but once the
bubble collapsed, that liquidity source dried up, and instead most
sellers were new entrants that had joined during the bubble. The steep
price decrease of 2022 Q1-Q2 led many of them to sell, driving prices
further down, thereby perpetuating the cycle.

\begin{figure}[htbp]
    \centering
    \includegraphics[width=0.9\columnwidth]{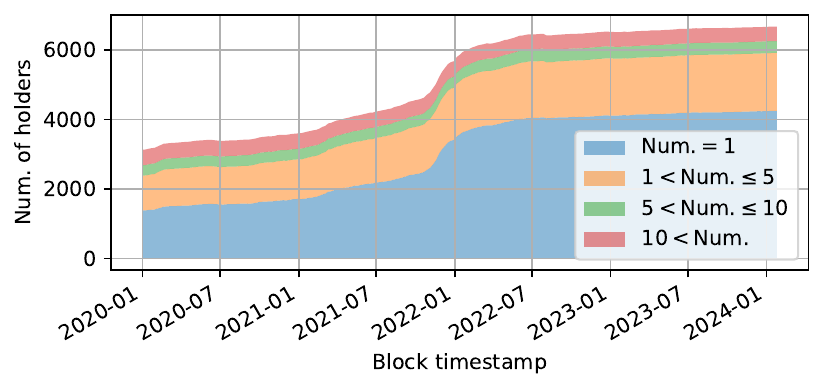}
    \caption{Unique landowners divided into tiers based on the number of parcels they own.
    The sum of the number of parcels held is always equal to the number of ownable parcels (i.e., 44,480).}
    \label{fig:num_parcel_holder}
\end{figure}

Figure~\ref{fig:num_parcel_holder} shows the number of unique landowners divided into tiers based on the number of parcels owned.
More than half of the landowners (early adopters) held multiple parcels at the beginning of 2020.
However, the number of single-parcel owners increased thereafter. In particular, we see an abrupt jump, of $\approx$1,000 single-parcel owners in 2021 Q4.
On the other hand, the number of owners who possess more than five parcels of land does not change in that period. In other words, holders of large number of parcels sold land to single-parcel holders, most likely new entrants to the market. 

In summary, the rapid appreciation of LAND tokens during the bubble
profited early adopters who sold (some of) their possessions then.
On the other hand, newly entrants who participated in
land sales, probably motivated by the surge in the popularity of the
metaverse, failed to reap profit from the sale and even frequently
incurred losses. We see a transfer of wealth from new participants
to early adopters---a situation reminiscent of what Soska~et~al.~\cite{Soska_Dong_Khodaverdian_Zetlin-Jones_Routledge_Christin_2021} had observed on BitMEX with larger portfolio holders continuously acquiring wealth from retail-level, less experienced investors.

\section{Conclusion}
\label{sec:dl_conclusion}
This study has explored transactions on Decentraland, a decentralized virtual world platform. 
LAND tokens, which represent digital property title, saw their price skyrocket in 2021, which correlated with an increased amount of online discussions about Decentraland, specifically on Reddit. 
Topic modeling analysis revealed that discussions about Decentraland 
viewed that platform as representative of the NFT and metaverse ecosystems, 
and frequently focused on strategies to profit from these ecosystems. 
As a result, it is likely that 
 Decentraland benefited from the hype surrounding NFTs and the metaverse.
We performed regression and profit-and-loss analyses of land trades. We
discovered that before the 2021 hype cycle, LAND prices followed established real-estate (bid-rent) theory, mimicking what we would see in the
physical world. During that hype cycle,  prices became
disconnected from geographical considerations, leading the market to be
subject to rampant short-term speculation---illustrated by the large amount of
parcels being ``flipped'' over short periods of time, and characteristic of an economic bubble. Large landowners,
with multiple parcels to their names, played an important role as both buyers and sellers at the same time.
In the aftermath of the bubble, new entrants were left with unrealized financial losses reaching hundreds of dollars, or were forced to engage in fire sales to cut off their losses.

Our analysis shows that early adopters,
who often had very large holdings, provided liquidity to the market and
made sizable profits (10,000-15,000 USD per parcel, give or take) by
selling some (but not all) of their holdings to new entrants that had
arrived during the bubble. On the other hand, these new entrants made little or no
profit -- and even frequently lost money in the order of 1,000 USD.

To summarize, we can describe the 2021 Decentraland boom, 
likely fueled by the rising popularity of the topics related to NFT and metaverse, as a
textbook case of bad investment by novice investors. Many new entrants
tried to cash in on the hype, and failed. 
Did these new entrants have sufficient
information to make sensible financial decisions? Platforms like
Decentraland do not resemble traditional financial (or real-estate)
investment platforms,
which can lead to investor confusion in terms of not
understanding the economic risks associated with them. As we demonstrated,
these risks are real: how to mitigate them is an avenue for future work,
that likely will include mandatory disclosures
and user education.

\ifanon
\else
  \section{Acknowledgements}
This research was partially supported by Ripple's University 
Block\-chain Research Initiative (UBRI) at Carnegie Mellon and by the 
Carnegie Mellon CyLab Secure Blockchain Initiative. 
D.K. was supported by the Japanese Government Long-Term Overseas Fellowship Program.
Some of the authors hold non-negligible cryptocurrency positions, but none on 
Decentraland. As part of exploratory efforts, some of the authors acquired a negligible amount of Decentraland real estate.

\fi


\bibliographystyle{informs2014}
\bibliography{main}

\begin{thebibliography}{62}
\providecommand{\natexlab}[1]{#1}
\providecommand{\url}[1]{\texttt{#1}}
\providecommand{\urlprefix}{URL }

\bibitem[{Adam et~al.(2017)Adam, Marcet, \protect\BIBand{}
  Beutel}]{Adam_Marcet_Beutel_2017}
Adam K, Marcet A, Beutel J (2017) Stock price booms and expected capital gains.
  \emph{American Economic Review} 107(8):2352–2408, ISSN 0002-8282,
  \urlprefix\url{http://dx.doi.org/10.1257/aer.20140205}.

\bibitem[{Aliber et~al.(2023)Aliber, Kindleberger, \protect\BIBand{}
  McCauley}]{Aliber_Kindleberger_McCauley_2023}
Aliber RZ, Kindleberger CP, McCauley RN (2023) \emph{Manias, Panics, and
  Crashes: A History of Financial Crises} (Cham: Springer International
  Publishing), ISBN 978-3-031-16007-3,
  \urlprefix\url{http://dx.doi.org/10.1007/978-3-031-16008-0}.

\bibitem[{Alonso(1960)}]{Alonso_1960}
Alonso W (1960) A theory of the urban land market. \emph{Papers in Regional
  Science} 6(1):149–157, ISSN 1435-5957,
  \urlprefix\url{http://dx.doi.org/10.1111/j.1435-5597.1960.tb01710.x}.

\bibitem[{Alonso(1964)}]{Alonso_1964}
Alonso W (1964) \emph{Location and Land Use: Toward a General Theory of Land
  Rent} (Cambridge, MA: Harvard University Press), ISBN 978-0-674-73085-4.

\bibitem[{Asriyan et~al.(2021)Asriyan, Fornaro, Martin, \protect\BIBand{}
  Ventura}]{Asriyan_Fornaro_Martin_Ventura_2021}
Asriyan V, Fornaro L, Martin A, Ventura J (2021) Monetary policy for a bubbly
  world. \emph{The Review of Economic Studies} 88(3):1418–1456, ISSN
  0034-6527, \urlprefix\url{http://dx.doi.org/10.1093/restud/rdaa045}.

\bibitem[{Auxier \protect\BIBand{} Anderson(2021)}]{Auxier_Anderson_2021}
Auxier B, Anderson M (2021) Social media use in 2021.
  \urlprefix\url{https://www.pewresearch.org/internet/2021/04/07/social-media-use-in-2021/}.

\bibitem[{Bakshy et~al.(2010)Bakshy, Simmons, Huffaker, Cheng,
  \protect\BIBand{} Adamic}]{Bakshy_Simmons_Huffaker_Cheng_Adamic_2010}
Bakshy E, Simmons M, Huffaker D, Cheng CY, Adamic L (2010) The social dynamics
  of economic activity in a virtual world. \emph{Proceedings of the
  International AAAI Conference on Web and Social Media} 4(11):2–9, ISSN
  2334-0770, \urlprefix\url{http://dx.doi.org/10.1609/icwsm.v4i1.14032}.

\bibitem[{Barberis et~al.(2015)Barberis, Greenwood, Jin, \protect\BIBand{}
  Shleifer}]{Barberis_Greenwood_Jin_Shleifer_2015}
Barberis N, Greenwood R, Jin L, Shleifer A (2015) X-capm: An extrapolative
  capital asset pricing model. \emph{Journal of Financial Economics}
  115(1):1–24, ISSN 0304-405X,
  \urlprefix\url{http://dx.doi.org/10.1016/j.jfineco.2014.08.007}.

\bibitem[{Barberis et~al.(2018)Barberis, Greenwood, Jin, \protect\BIBand{}
  Shleifer}]{Barberis_Greenwood_Jin_Shleifer_2018}
Barberis N, Greenwood R, Jin L, Shleifer A (2018) Extrapolation and bubbles.
  \emph{Journal of Financial Economics} 129(2):203–227, ISSN 0304-405X,
  \urlprefix\url{http://dx.doi.org/10.1016/j.jfineco.2018.04.007}.

\bibitem[{Baumgartner et~al.(2020)Baumgartner, Zannettou, Keegan, Squire,
  \protect\BIBand{}
  Blackburn}]{Baumgartner_Zannettou_Keegan_Squire_Blackburn_2020}
Baumgartner J, Zannettou S, Keegan B, Squire M, Blackburn J (2020) The
  pushshift reddit dataset. \emph{Proceedings of the International AAAI
  Conference on Web and Social Media} 14:830–839, ISSN 2334-0770,
  \urlprefix\url{http://dx.doi.org/10.1609/icwsm.v14i1.7347}.

\bibitem[{Baur et~al.(2018)Baur, Hong, \protect\BIBand{}
  Lee}]{Baur_Hong_Lee_2018}
Baur DG, Hong K, Lee AD (2018) Bitcoin: Medium of exchange or speculative
  assets? \emph{Journal of International Financial Markets, Institutions and
  Money} 54:177–189,
  \urlprefix\url{http://dx.doi.org/10.1016/j.intfin.2017.12.004}.

\bibitem[{Biais et~al.(2023)Biais, Capponi, Cong, Gaur, \protect\BIBand{}
  Giesecke}]{Biais_Capponi_Cong_Gaur_Giesecke_2023}
Biais B, Capponi A, Cong LW, Gaur V, Giesecke K (2023) Advances in blockchain
  and crypto economics. \emph{Management Science} 69(11):6417–6426, ISSN
  0025-1909, \urlprefix\url{http://dx.doi.org/10.1287/mnsc.2023.intro.v69.n11}.

\bibitem[{Boellstorff(2015)}]{Boellstorff_2015}
Boellstorff T (2015) \emph{Coming of Age in Second Life: An Anthropologist
  Explores the Virtually Human} (Princeton University Press), ISBN
  978-1-4008-7410-1, \urlprefix\url{http://dx.doi.org/10.1515/9781400874101}.

\bibitem[{Bordalo et~al.(2018)Bordalo, Gennaioli, \protect\BIBand{}
  Shleifer}]{Bordalo_Gennaioli_Shleifer_2018}
Bordalo P, Gennaioli N, Shleifer A (2018) Diagnostic expectations and credit
  cycles. \emph{The Journal of Finance} 73(1):199–227, ISSN 1540-6261,
  \urlprefix\url{http://dx.doi.org/10.1111/jofi.12586}.

\bibitem[{Case et~al.(2012)Case, Shiller, \protect\BIBand{}
  Thompson}]{Case_Shiller_Thompson_2012}
Case KE, Shiller RJ, Thompson A (2012) What have they been thinking? home buyer
  behavior in hot and cold markets (18400),
  \urlprefix\url{http://dx.doi.org/10.3386/w18400}, dOI: 10.3386/w18400.

\bibitem[{Cassella \protect\BIBand{} Gulen(2018)}]{Cassella_Gulen_2018}
Cassella S, Gulen H (2018) Extrapolation bias and the predictability of stock
  returns by price-scaled variables. \emph{The Review of Financial Studies}
  31(11):4345–4397, ISSN 0893-9454,
  \urlprefix\url{http://dx.doi.org/10.1093/rfs/hhx139}.

\bibitem[{Cheng et~al.(2019)Cheng, De~Franco, Jiang, \protect\BIBand{}
  Lin}]{Cheng_Franco_Jiang_Lin_2019}
Cheng SF, De~Franco G, Jiang H, Lin P (2019) Riding the blockchain mania:
  Public firms’ speculative 8-k disclosures. \emph{Management Science}
  65(12):5901–5913, ISSN 0025-1909,
  \urlprefix\url{http://dx.doi.org/10.1287/mnsc.2019.3357}.

\bibitem[{Christie’s(2021)}]{Christies_Beeple}
Christie’s (2021) Beeple: A visionary digital artist at the forefront of
  nfts.
  \urlprefix\url{https://www.christies.com/en/stories/monumental-collage-by-beeple-is-first-purely-digital-artwork-nft-to-come-to-auction-0463a2c0f3174b17997fba8a1fe4c865}.

\bibitem[{Cooper et~al.(2001)Cooper, Dimitrov, \protect\BIBand{}
  Rau}]{Cooper_Dimitrov_Rau_2001}
Cooper MJ, Dimitrov O, Rau PR (2001) A rose.com by any other name. \emph{The
  Journal of Finance} 56(6):2371–2388, ISSN 1540-6261,
  \urlprefix\url{http://dx.doi.org/10.1111/0022-1082.00408}.

\bibitem[{Cutler et~al.(1990)Cutler, Poterba, \protect\BIBand{}
  Summers}]{Cutler_Poterba_Summers_1990}
Cutler DM, Poterba JM, Summers LH (1990) Speculative dynamics and the role of
  feedback traders (3243), \urlprefix\url{http://dx.doi.org/10.3386/w3243},
  dOI: 10.3386/w3243.

\bibitem[{De~Long et~al.(1990)De~Long, Shleifer, Summers, \protect\BIBand{}
  Waldmann}]{De_Long_Shleifer_Summers_Waldmann_1990}
De~Long JB, Shleifer A, Summers LH, Waldmann RJ (1990) Positive feedback
  investment strategies and destabilizing rational speculation. \emph{The
  Journal of Finance} 45(2):379–395, ISSN 1540-6261,
  \urlprefix\url{http://dx.doi.org/10.1111/j.1540-6261.1990.tb03695.x}.

\bibitem[{Decentraland(2021)}]{Decentraland_2021_gallery}
Decentraland (2021) Sotheby’s opens a virtual gallery in decentraland.
  \urlprefix\url{https://decentraland.org/blog/announcements/sotheby-s-opens-a-virtual-gallery-in-decentraland}.

\bibitem[{DeFusco et~al.(2017)DeFusco, Nathanson, \protect\BIBand{}
  Zwick}]{DeFusco_Nathanson_Zwick_2017}
DeFusco AA, Nathanson CG, Zwick E (2017) Speculative dynamics of prices and
  volume (23449), \urlprefix\url{http://dx.doi.org/10.3386/w23449}, dOI:
  10.3386/w23449.

\bibitem[{Entriken et~al.(2018)Entriken, Shirley, Evans, \protect\BIBand{}
  Sachs}]{Entriken_Shirley_Evans_Sachs_2018}
Entriken W, Shirley D, Evans J, Sachs N (2018) Erc-721: Non-fungible token
  standard. From: https://eips.ethereum.org/EIPS/eip-721.

\bibitem[{{European Council}(2023)}]{EU_2023}
{European Council} (2023) Digital finance: Council adopts new rules on markets
  in crypto-assets (mica). From:
  https://www.consilium.europa.eu/en/press/press-releases/2023/05/16/digital-finance-council-adopts-new-rules-on-markets-in-crypto-assets-mica/.

\bibitem[{Faverio \protect\BIBand{} Sidoti(2023)}]{Faverio_Sidoti_2023}
Faverio M, Sidoti O (2023) Majority of americans aren’t confident in the
  safety and reliability of cryptocurrency.
  \urlprefix\url{https://www.pewresearch.org/short-reads/2023/04/10/majority-of-americans-arent-confident-in-the-safety-and-reliability-of-cryptocurrency/}.

\bibitem[{Fujita \protect\BIBand{} Thisse(2002)}]{Fujita_Thisse_2002}
Fujita M, Thisse JF (2002) \emph{Economics of Agglomeration: Cities, Industrial
  Location, and Regional Growth} (Cambridge: Cambridge University Press),
  \urlprefix\url{http://dx.doi.org/10.1017/CBO9780511805660}.

\bibitem[{Gennaioli et~al.(2012)Gennaioli, Shleifer, \protect\BIBand{}
  Vishny}]{Gennaioli_Shleifer_Vishny_2012}
Gennaioli N, Shleifer A, Vishny R (2012) Neglected risks, financial innovation,
  and financial fragility. \emph{Journal of Financial Economics}
  104(3):452–468, ISSN 0304-405X,
  \urlprefix\url{http://dx.doi.org/10.1016/j.jfineco.2011.05.005}.

\bibitem[{Gertler et~al.(2020)Gertler, Kiyotaki, \protect\BIBand{}
  Prestipino}]{Gertler_Kiyotaki_Prestipino_2020}
Gertler M, Kiyotaki N, Prestipino A (2020) A macroeconomic model with financial
  panics. \emph{The Review of Economic Studies} 87(1):240–288, ISSN
  0034-6527, \urlprefix\url{http://dx.doi.org/10.1093/restud/rdz032}.

\bibitem[{Glaeser \protect\BIBand{} Nathanson(2017)}]{Glaeser_Nathanson_2017}
Glaeser EL, Nathanson CG (2017) An extrapolative model of house price dynamics.
  \emph{Journal of Financial Economics} 126(1):147–170, ISSN 0304-405X,
  \urlprefix\url{http://dx.doi.org/10.1016/j.jfineco.2017.06.012}.

\bibitem[{Gupta et~al.(2022)Gupta, Mittal, Peeters, \protect\BIBand{}
  Van~Nieuwerburgh}]{Gupta_Mittal_Peeters_Van_Nieuwerburgh_2022}
Gupta A, Mittal V, Peeters J, Van~Nieuwerburgh S (2022) Flattening the curve:
  Pandemic-induced revaluation of urban real estate. \emph{Journal of Financial
  Economics} 146(2):594–636, ISSN 0304-405X,
  \urlprefix\url{http://dx.doi.org/10.1016/j.jfineco.2021.10.008}.

\bibitem[{Hong \protect\BIBand{} Stein(1999)}]{Hong_Stein_1999}
Hong H, Stein JC (1999) A unified theory of underreaction, momentum trading,
  and overreaction in asset markets. \emph{The Journal of Finance}
  54(6):2143–2184, ISSN 1540-6261,
  \urlprefix\url{http://dx.doi.org/10.1111/0022-1082.00184}.

\bibitem[{Hong \protect\BIBand{} Stein(2007)}]{Hong_Stein_2007}
Hong H, Stein JC (2007) Disagreement and the stock market. \emph{Journal of
  Economic Perspectives} 21(2):109–128, ISSN 0895-3309,
  \urlprefix\url{http://dx.doi.org/10.1257/jep.21.2.109}.

\bibitem[{Jhaver et~al.(2019)Jhaver, Bruckman, \protect\BIBand{}
  Gilbert}]{Jhaver_Bruckman_Gilbert_2019}
Jhaver S, Bruckman A, Gilbert E (2019) Does transparency in moderation really
  matter? user behavior after content removal explanations on reddit.
  \emph{Proceedings of the ACM on Human-Computer Interaction}
  3(CSCW):150:1--150:27, \urlprefix\url{http://dx.doi.org/10.1145/3359252}.

\bibitem[{Jin \protect\BIBand{} Sui(2022)}]{Jin_Sui_2022}
Jin LJ, Sui P (2022) Asset pricing with return extrapolation. \emph{Journal of
  Financial Economics} 145(2, Part A):273–295, ISSN 0304-405X,
  \urlprefix\url{http://dx.doi.org/10.1016/j.jfineco.2021.10.009}.

\bibitem[{Joshua(2017)}]{Joshua_2017}
Joshua J (2017) Information bodies: Computational anxiety in neal
  stephenson’s snow crash. \emph{Interdisciplinary Literary Studies}
  19(1):17–47,
  \urlprefix\url{http://dx.doi.org/10.5325/intelitestud.19.1.0017}, publisher:
  Penn State University Press.

\bibitem[{Junqu\'{e}~de Fortuny \protect\BIBand{}
  Zhang(2023)}]{Junque_Zhang_2023}
Junqu\'{e}~de Fortuny E, Zhang Y (2023) Exploring the new frontier:
  Decentralized financial services. \emph{Service Science} 15(4):266–282,
  ISSN 2164-3962, \urlprefix\url{http://dx.doi.org/10.1287/serv.2021.0048}.

\bibitem[{Kahneman \protect\BIBand{} Tversky(1979)}]{Kahneman_Tversky_1979}
Kahneman D, Tversky A (1979) Prospect theory: An analysis of decision under
  risk. \emph{Econometrica} 47(2):263–291,
  \urlprefix\url{http://dx.doi.org/10.2307/1914185}.

\bibitem[{Kawai et~al.(2023)Kawai, Cuevas, Routledge, Soska, Zetlin-Jones,
  \protect\BIBand{} Christin}]{dkawai_2023}
Kawai D, Cuevas A, Routledge B, Soska K, Zetlin-Jones A, Christin N (2023) Is
  your digital neighbor a reliable investment advisor? \emph{Proceedings of the
  ACM Web Conference 2023}, 3581–3591, WWW '23 (New York, NY, USA:
  Association for Computing Machinery), ISBN 9781450394161,
  \urlprefix\url{http://dx.doi.org/10.1145/3543507.3583502}.

\bibitem[{Kumar et~al.(2022)Kumar, McLaughlim, Xie, Nicolet-Serra, Müller,
  \protect\BIBand{} Rigg}]{Kumar_McLaughlim_Xie_Nicolet-Serra_Muller_Rigg}
Kumar SJ, McLaughlim JM, Xie AL, Nicolet-Serra L, Müller A, Rigg C (2022) The
  nft collection: A brave nft world – a regulatory review of nfts (part 2).
  \urlprefix\url{https://www.natlawreview.com/article/nft-collection-brave-nft-world-regulatory-review-nfts-part-2}.

\bibitem[{Lee et~al.(2021)Lee, Braud, Zhou, Wang, Xu, Lin, Kumar, Bermejo,
  \protect\BIBand{} Hui}]{Lee_Braud_Zhou_Wang_Xu_Lin_Kumar_Bermejo_Hui_2021}
Lee LH, Braud T, Zhou P, Wang L, Xu D, Lin Z, Kumar A, Bermejo C, Hui P (2021)
  All one needs to know about metaverse: A complete survey on technological
  singularity, virtual ecosystem, and research agenda (arXiv:2110.05352),
  \urlprefix\url{http://dx.doi.org/10.48550/arXiv.2110.05352}, arXiv:2110.05352
  [cs].

\bibitem[{Liu et~al.(2015)Liu, Zhang, \protect\BIBand{}
  Zhao}]{Liu_Zhang_Zhao_2015}
Liu YJ, Zhang Z, Zhao L (2015) Speculation spillovers. \emph{Management
  Science} 61(3):649–664, ISSN 0025-1909,
  \urlprefix\url{http://dx.doi.org/10.1287/mnsc.2014.1914}.

\bibitem[{Maxted(2023)}]{Maxted_2023}
Maxted P (2023) A macro-finance model with sentiment. \emph{The Review of
  Economic Studies} rdad023, ISSN 0034-6527,
  \urlprefix\url{http://dx.doi.org/10.1093/restud/rdad023}.

\bibitem[{Messinger et~al.(2009)Messinger, Stroulia, Lyons, Bone, Niu, Smirnov,
  \protect\BIBand{}
  Perelgut}]{Messinger_Stroulia_Lyons_Bone_Niu_Smirnov_Perelgut_2009}
Messinger PR, Stroulia E, Lyons K, Bone M, Niu RH, Smirnov K, Perelgut S (2009)
  Virtual worlds — past, present, and future: New directions in social
  computing. \emph{Decision Support Systems} 47(3):204–228, ISSN 0167-9236,
  \urlprefix\url{http://dx.doi.org/10.1016/j.dss.2009.02.014}.

\bibitem[{Meta(2021)}]{Meta_2021}
Meta (2021) The facebook company is now meta.
  \urlprefix\url{https://about.fb.com/news/2021/10/facebook-company-is-now-meta/}.

\bibitem[{Minsky(1977)}]{Minsky_1977}
Minsky HP (1977) The financial instability hypothesis: An interpretation of
  keynes and an alternative to “standard” theory. \emph{Challenge}
  20(1):20–27, ISSN 0577-5132.

\bibitem[{Mystakidis(2022)}]{Mystakidis_2022}
Mystakidis S (2022) Metaverse. \emph{Encyclopedia} 2(1):486–497,
  \urlprefix\url{http://dx.doi.org/10.3390/encyclopedia2010031}.

\bibitem[{Nadini et~al.(2021)Nadini, Alessandretti, Di~Giacinto, Martino,
  Aiello, \protect\BIBand{}
  Baronchelli}]{Nadini_Alessandretti_Di_Giacinto_Martino_Aiello_Baronchelli_2021}
Nadini M, Alessandretti L, Di~Giacinto F, Martino M, Aiello LM, Baronchelli A
  (2021) Mapping the nft revolution: market trends, trade networks, and visual
  features. \emph{Scientific Reports} 11(11):20902, ISSN 2045-2322,
  \urlprefix\url{http://dx.doi.org/10.1038/s41598-021-00053-8}.

\bibitem[{Nakamoto(2008)}]{Nakamoto_2008}
Nakamoto S (2008) Bitcoin: A peer-to-peer electronic cash system
  \urlprefix\url{https://bitcoin.org/bitcoin.pdf}.

\bibitem[{Ofek \protect\BIBand{} Richardson(2003)}]{Ofek_Richardson_2003}
Ofek E, Richardson M (2003) Dotcom mania: The rise and fall of internet stock
  prices. \emph{The Journal of Finance} 58(3):1113–1137, ISSN 1540-6261,
  \urlprefix\url{http://dx.doi.org/10.1111/1540-6261.00560}.

\bibitem[{Ordano et~al.(2017)Ordano, Meilich, Jardi, \protect\BIBand{}
  Araoz}]{decentraland_wp}
Ordano E, Meilich A, Jardi Y, Araoz M (2017) Decentraland a blockchain-based
  virtual world. \urlprefix\url{https://decentraland.org/whitepaper.pdf}.

\bibitem[{P\'{e}nasse \protect\BIBand{}
  Renneboog(2022)}]{Penasse_Renneboog_2022}
P\'{e}nasse J, Renneboog L (2022) Speculative trading and bubbles: Evidence
  from the art market. \emph{Management Science} 68(7):4939–4963, ISSN
  0025-1909, \urlprefix\url{http://dx.doi.org/10.1287/mnsc.2021.4088}.

\bibitem[{Proferes et~al.(2021)Proferes, Jones, Gilbert, Fiesler,
  \protect\BIBand{} Zimmer}]{Proferes_Jones_Gilbert_Fiesler_Zimmer_2021}
Proferes N, Jones N, Gilbert S, Fiesler C, Zimmer M (2021) Studying reddit: A
  systematic overview of disciplines, approaches, methods, and ethics.
  \emph{Social Media + Society} 7(2):20563051211019004, ISSN 2056-3051,
  \urlprefix\url{http://dx.doi.org/10.1177/20563051211019004}.

\bibitem[{Qiang et~al.(2022)Qiang, Qian, Li, Yuan, \protect\BIBand{}
  Wu}]{Qiang_Qian_Li_Yuan_Wu_2022}
Qiang J, Qian Z, Li Y, Yuan Y, Wu X (2022) Short text topic modeling
  techniques, applications, and performance: A survey. \emph{IEEE Transactions
  on Knowledge and Data Engineering} 34(03):1427–1445, ISSN 1041-4347,
  \urlprefix\url{http://dx.doi.org/10.1109/TKDE.2020.2992485}.

\bibitem[{R\"{o}der et~al.(2015)R\"{o}der, Both, \protect\BIBand{}
  Hinneburg}]{Roder_Both_Hinneburg_2015}
R\"{o}der M, Both A, Hinneburg A (2015) Exploring the space of topic coherence
  measures. \emph{Proceedings of the Eighth ACM International Conference on Web
  Search and Data Mining}, 399–408, WSDM ’15 (New York, NY, USA:
  Association for Computing Machinery), ISBN 978-1-4503-3317-7,
  \urlprefix\url{http://dx.doi.org/10.1145/2684822.2685324}.

\bibitem[{{Securities and Exchange Commission}(2023)}]{SEC_NFT}
{Securities and Exchange Commission} (2023) Framework for “investment
  contract” analysis of digital assets.
  \urlprefix\url{https://www.sec.gov/corpfin/framework-investment-contract-analysis-digital-assets#}.

\bibitem[{Shelton(2010)}]{Shelton_2010}
Shelton AK (2010) Defining the lines between virtual and real world purchases:
  Second life sells, but who’s buying? \emph{Computers in Human Behavior}
  26(6):1223–1227,
  \urlprefix\url{http://dx.doi.org/10.1016/j.chb.2010.03.019}.

\bibitem[{Soska et~al.(2021)Soska, Dong, Khodaverdian, Zetlin-Jones, Routledge,
  \protect\BIBand{}
  Christin}]{Soska_Dong_Khodaverdian_Zetlin-Jones_Routledge_Christin_2021}
Soska K, Dong JD, Khodaverdian A, Zetlin-Jones A, Routledge B, Christin N
  (2021) Towards understanding cryptocurrency derivatives:a case study of
  bitmex. \emph{Proceedings of the Web Conference 2021}, 45–57, WWW ’21
  (Association for Computing Machinery), ISBN 9781450383127,
  \urlprefix\url{http://dx.doi.org/10.1145/3442381.3450059}.

\bibitem[{Stokes(2012)}]{Stokes_2012}
Stokes R (2012) Virtual money laundering: the case of bitcoin and the linden
  dollar. \emph{Information \& Communications Technology Law} 21(3):221–236,
  ISSN 1360-0834,
  \urlprefix\url{http://dx.doi.org/10.1080/13600834.2012.744225}.

\bibitem[{Torres(2023)}]{SJVB2023}
Torres D (2023)
  \url{https://www.sfvbj.com/technology/getting-children-hooked-on-nfts}. Last
  accessed: \today.

\bibitem[{Weisberg(2005)}]{Weisberg_2005}
Weisberg S (2005) \emph{Outliers and Influence}, 194–210 (John Wiley \& Sons,
  Ltd), ISBN 978-0-471-70409-6,
  \urlprefix\url{http://dx.doi.org/10.1002/0471704091.ch9}.

\bibitem[{Yin \protect\BIBand{} Wang(2014)}]{Yin_Wang_2014}
Yin J, Wang J (2014) A dirichlet multinomial mixture model-based approach for
  short text clustering. \emph{Proceedings of the 20th ACM SIGKDD international
  conference on Knowledge discovery and data mining} 233–242,
  \urlprefix\url{http://dx.doi.org/10.1145/2623330.2623715}.

\end{thebibliography}

\clearpage

\appendix
\section{Supplementary information about Reddit analysis}
\label{appx:reddit_exclusion}

\paragraph{Bot-like message in Reddit submission posts}
This section discusses bot-like submissions on Reddit.

Figure~\ref{fig:num_posts} shows a significant increase in the number of submissions in Oct. 2021.
We address the cause of this increase in this section.
Table~\ref{tab:tf_idf_title_compare} summarizes the top five words on the titles of Reddit submissions ranked by TF-IDF score, where TF-IDF scores for the $i$-th period are calculated as follows:
$$
V_{w, i} = TF_{w,i} \times \log \left(\frac{\left| \mathbf{D} \right|}{DF_{w}} \right) \, ,
$$
where $TF_{w,i}$ and $DF_{w}$ are the term frequency of a word $w$ in the $i$-th observation period and document frequency of the word $w$, respectively.
We normalize $V_{w,i}$ over words for each period: $\hat{V}_{w,i} \equiv V_{w,i} / \sqrt{\sum_w V_{w,i}^2}$.

\begin{table}[htbp]
    \caption{Top five words based on TF-IDF analysis for the titles of submissions with ``Decentraland.'' The values in parentheses represent the TF-IDF scores for words $\hat{V}_{w,i}$.
    The table on the left and that on the right summarize TF-IDF results for submissions referring to ``Decentraland'' and those excluding ``coinsniper,'' respectively.}
    \label{tab:tf_idf_title_compare}
\begin{minipage}{0.44\textwidth}
    \centering
    \resizebox{0.95\columnwidth}{!}{
    \addtolength{\tabcolsep}{-4.2pt}    
    \begin{tabular}{cccccc} \toprule
        &\multicolumn{5}{c}{TF-IDF for submissions w/ ``Decentraland''} 
        \\ \cmidrule{2-6}
        & 
        1st & 2nd & 3rd & 4th & 5th
        \\\midrule
        \begin{tabular}{c}
        '19 Q1-'20 Q4
        \end{tabular}
        & 
        \begin{tabular}{c}virtual\\(0.311)\end{tabular}&
        \begin{tabular}{c}game\\(0.248)\end{tabular}&
        \begin{tabular}{c}ethereum\\(0.242)\end{tabular}&
        \begin{tabular}{c}world\\(0.220)\end{tabular}&
        \begin{tabular}{c}launch\\(0.198)\end{tabular}
        \\
        \begin{tabular}{c}
        '21 Q1
        \end{tabular}
        &
        \begin{tabular}{c}game\\(0.275)\end{tabular}&
        \begin{tabular}{c}pizza\\(0.233)\end{tabular}&
        \begin{tabular}{c}virtual\\(0.231)\end{tabular}&
        \begin{tabular}{c}ethereum\\(0.194)\end{tabular}&
        \begin{tabular}{c}atari\\(0.189)\end{tabular}
        \\
        \begin{tabular}{c}
        '21 Q2
        \end{tabular}
        &
        \begin{tabular}{c}poocoin\\(0.523)\end{tabular}&
        \begin{tabular}{c}koing\\(0.311)\end{tabular}&
        \begin{tabular}{c}pre\\(0.261)\end{tabular}&
        \begin{tabular}{c}sale\\(0.214)\end{tabular}&
        \begin{tabular}{c}ad\\(0.200)\end{tabular}
        \\
        \begin{tabular}{c}
        '21 Q3
        \end{tabular}
        &
        \begin{tabular}{c}ga\\(0.439)\end{tabular}&
        \begin{tabular}{c}land\\(0.278)\end{tabular}&
        \begin{tabular}{c}decentralize\\(0.258)\end{tabular}&
        \begin{tabular}{c}property\\(0.258)\end{tabular}&
        \begin{tabular}{c}sell\\(0.247)\end{tabular}
        \\
        \begin{tabular}{c}
        '21 Q4
        \end{tabular}
        &
        \begin{tabular}{c}\textbf{coinsniper}\\(0.403)\end{tabular}&
        \begin{tabular}{c}lock\\(0.342)\end{tabular}&
        \begin{tabular}{c}later\\(0.340)\end{tabular}&
        \begin{tabular}{c}mc\\(0.340)\end{tabular}&
        \begin{tabular}{c}lp\\(0.290)\end{tabular}
        \\
        \begin{tabular}{c}
        '22 Q1
        \end{tabular}
        &
        \begin{tabular}{c}metaverse\\(0.410)\end{tabular}&
        \begin{tabular}{c}crypto\\(0.289)\end{tabular}&
        \begin{tabular}{c}today\\(0.259)\end{tabular}&
        \begin{tabular}{c}far\\(0.223)\end{tabular}&
        \begin{tabular}{c}trend\\(0.222)\end{tabular}
        \\
        \begin{tabular}{c}
        '22 Q2
        \end{tabular}
        &
        \begin{tabular}{c}metaverse\\(0.640)\end{tabular}&
        \begin{tabular}{c}bsc\\(0.201)\end{tabular}&
        \begin{tabular}{c}tipsycoin\\(0.201)\end{tabular}&
        \begin{tabular}{c}minute\\(0.201)\end{tabular}&
        \begin{tabular}{c}project\\(0.179)\end{tabular}
        \\
        \begin{tabular}{c}
        '22 Q3
        \end{tabular}
        &
        \begin{tabular}{c}metaverse\\(0.542)\end{tabular}&
        \begin{tabular}{c}crypto\\(0.369)\end{tabular}&
        \begin{tabular}{c}today\\(0.235)\end{tabular}&
        \begin{tabular}{c}trend\\(0.232)\end{tabular}&
        \begin{tabular}{c}far\\(0.227)\end{tabular}
        \\
        \begin{tabular}{c}
        '22 Q4
        \end{tabular}
        &
        \begin{tabular}{c}crypto\\(0.460)\end{tabular}&
        \begin{tabular}{c}today\\(0.454)\end{tabular}&
        \begin{tabular}{c}trend\\(0.401)\end{tabular}&
        \begin{tabular}{c}far\\(0.397)\end{tabular}&
        \begin{tabular}{c}metaverse\\(0.273)\end{tabular}
        \\\bottomrule
    \end{tabular} 
    \addtolength{\tabcolsep}{4.2pt}    
    }
\end{minipage}
\begin{minipage}{0.51\textwidth}
    \centering
    \resizebox{\columnwidth}{!}{
    \addtolength{\tabcolsep}{-4.2pt}    
    \begin{tabular}{cccccc} \toprule
        &\multicolumn{5}{c}{TF-IDF for submissions w/ ''Decentraland'' excl. those w/ ``coinsniper''} 
        \\ \cmidrule{2-6}
        & 
        1st & 2nd & 3rd & 4th & 5th
        \\\midrule
        \begin{tabular}{c}
        '19 Q1-'20 Q4
        \end{tabular}
        & 
        \begin{tabular}{c}virtual\\(0.309)\end{tabular}&
        \begin{tabular}{c}game\\(0.246)\end{tabular}&
        \begin{tabular}{c}ethereum\\(0.241)\end{tabular}&
        \begin{tabular}{c}world\\(0.219)\end{tabular}&
        \begin{tabular}{c}launch\\(0.197)\end{tabular}
        \\
        \begin{tabular}{c}
        '21 Q1
        \end{tabular}
        &
        \begin{tabular}{c}game\\(0.269)\end{tabular}&
        \begin{tabular}{c}pizza\\(0.258)\end{tabular}&
        \begin{tabular}{c}virtual\\(0.226)\end{tabular}&
        \begin{tabular}{c}atari\\(0.205)\end{tabular}&
        \begin{tabular}{c}casino\\(0.196)\end{tabular}
        \\
        \begin{tabular}{c}
        '21 Q2
        \end{tabular}
        &
        \begin{tabular}{c}poocoin\\(0.578)\end{tabular}&
        \begin{tabular}{c}koing\\(0.290)\end{tabular}&
        \begin{tabular}{c}pre\\(0.271)\end{tabular}&
        \begin{tabular}{c}sale\\(0.199)\end{tabular}&
        \begin{tabular}{c}dynamic\\(0.197)\end{tabular}
        \\
        \begin{tabular}{c}
        '21 Q3
        \end{tabular}
        &
        \begin{tabular}{c}ga\\(0.489)\end{tabular}&
        \begin{tabular}{c}land\\(0.269)\end{tabular}&
        \begin{tabular}{c}decentralize\\(0.249)\end{tabular}&
        \begin{tabular}{c}property\\(0.248)\end{tabular}&
        \begin{tabular}{c}sell\\(0.239)\end{tabular}
        \\
        \begin{tabular}{c}
        '21 Q4
        \end{tabular}
        &
        \begin{tabular}{c}\textbf{metaverse}\\(0.559)\end{tabular}&
        \begin{tabular}{c}minute\\(0.284)\end{tabular}&
        \begin{tabular}{c}cryptocurrency\\(0.204)\end{tabular}&
        \begin{tabular}{c}crypto\\(0.171)\end{tabular}&
        \begin{tabular}{c}price\\(0.145)\end{tabular}
        \\
        \begin{tabular}{c}
        '22 Q1
        \end{tabular}
        &
        \begin{tabular}{c}\textbf{metaverse}\\(0.410)\end{tabular}&
        \begin{tabular}{c}crypto\\(0.288)\end{tabular}&
        \begin{tabular}{c}today\\(0.259)\end{tabular}&
        \begin{tabular}{c}far\\(0.222)\end{tabular}&
        \begin{tabular}{c}trend\\(0.222)\end{tabular}
        \\
        \begin{tabular}{c}
        '22 Q2
        \end{tabular}
        &
        \begin{tabular}{c}\textbf{metaverse}\\(0.635)\end{tabular}&
        \begin{tabular}{c}bsc\\(0.200)\end{tabular}&
        \begin{tabular}{c}tipsycoin\\(0.200)\end{tabular}&
        \begin{tabular}{c}minute\\(0.200)\end{tabular}&
        \begin{tabular}{c}project\\(0.178)\end{tabular}
        \\
        \begin{tabular}{c}
        '22 Q3
        \end{tabular}
        &
        \begin{tabular}{c}\textbf{metaverse}\\(0.540)\end{tabular}&
        \begin{tabular}{c}crypto\\(0.368)\end{tabular}&
        \begin{tabular}{c}today\\(0.234)\end{tabular}&
        \begin{tabular}{c}trend\\(0.231)\end{tabular}&
        \begin{tabular}{c}far\\(0.227)\end{tabular}
        \\
        \begin{tabular}{c}
        '22 Q4
        \end{tabular}
        &
        \begin{tabular}{c}crypto\\(0.459)\end{tabular}&
        \begin{tabular}{c}today\\(0.453)\end{tabular}&
        \begin{tabular}{c}trend\\(0.400)\end{tabular}&
        \begin{tabular}{c}far\\(0.396)\end{tabular}&
        \begin{tabular}{c}\textbf{metaverse}\\(0.272)\end{tabular}
        \\\bottomrule
    \end{tabular} 
    \addtolength{\tabcolsep}{4.2pt}    
    }
\end{minipage}
\end{table}
The table on the left summarizes the results for submissions that refer to ``Decentraland.'' It shows that "coinsniper" is ranked as the top result, followed by other seemingly unrelated words.
We found that all the top five words in 4Q 2021 come from bot-like posts saying the same message. In fact, 1,350 out of 1,788 posts (75.5\%) submitted in the week of Oct.~24, 2021, are a single message that recommends investment in Decentraland-related cryptocurrencies and NFT collections in CoinSniper, a cryptocurrency investment platform featuring newly emergent cryptocurrencies, reflecting the surging price of MANA token and parcels in Decentraland.
Moreover, 1,989 posts in that quarter had the same message, which indicates bot activity.
The table on the right summarizes the TF-IDF result for the submissions referring to ``Decentraland,' excluding those referencing ``coinsniper'' too. It appears that ``metaverse'' has the highest rank, followed by other relevant words. This seems to be a reasonable result. Therefore, we omit the submissions referring to ``coinsniper'' from GSDMM analysis.

\paragraph{Details of DMM analysis}
This section describes our GSDMM-based analysis of the Reddit posts dataset. We first collect all submissions referring to ``Decentraland'' either in their titles or main bodies (See Figure~\ref{fig:reddit_diagram}).
We excluded posts with titles referencing "coinsniper" as they seemed like bot-generated posts (See the previous paragraph) and, then, built a corpus from their titles (average word count: 13.4). 
The number of submissions referring to ``Decentraland'' is 24,893, and 22,872 submissions out of them are included in our corpus due to the exclusion of those referring to ``coinsniper.''
For the corpus, we search for frequent bigrams and trigrams. 
Then, we lemmatize words and exclude English stopwords in the corpus. 
We conducted GSDMM analysis on the pre-processed corpus. 
To determine the number of topics in our model, we executed GSDMM calculation with varying numbers of topics under the condition of $\alpha=\beta = 0.1$.\footnote{See Yin~and~Wang~\cite{Yin_Wang_2014} for the definition of $\alpha$ and $\beta$.}
We opted for ten topics as that number yields the highest coherence score \cite{Roder_Both_Hinneburg_2015}.

\section{Supplementary information about regression analysis}
\label{appx:reg_suppl}
This section describes supplementary information on our regression analysis.

\paragraph{Distance dependence of the number of visitors to parcels}
This appendix considers the dependence of the number of visitors to parcels on the distance from the center of Decentraland map.
Decentraland provides the number of visitors to parcels over past 30 days.
We collect the data for the period from Apr.~15, 2024 to June~5, 2024.

\begin{figure}[htb]
    \centering
    \includegraphics[width=0.8\textwidth]{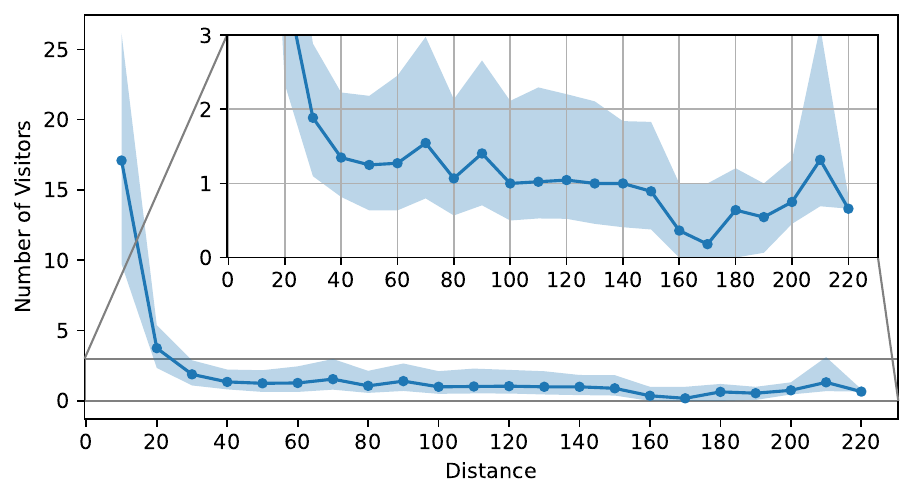}
    \caption{The median number of parcel visitors for the period from Apr.~15,~2024 to June~5, 2024.
    To calculate the median numbers, we first find the average number of visitors for each parcel over time. 
    Then, we calculate the median number of visitors for the parcels grouped by distance in 10-unit blocks.
    The area colored light blue represents the 25th to 75th percentile range for the blocks.
    }
    \label{fig:number_of_visitor}
\end{figure}

Figure~\ref{fig:number_of_visitor} shows the median number of parcel visitors.
To calculate the median numbers, we first find the average number of visitors for each parcel over time. 
Then, we calculate the median number of visitors for the parcels grouped by distance in 10-unit blocks. 
The result shows a decrease in the number of visitors as the distance from the center increases.
While we cannot directly assess whether this negative relationship holds true for the period from 2020 to 2023, this result suggests that the number of visitors negatively depends on the distance from the center as we hypothesize in Section~\ref{sec:dl_method}.

\begin{figure}[htb]
    \centering
    \includegraphics[width=0.8\columnwidth]{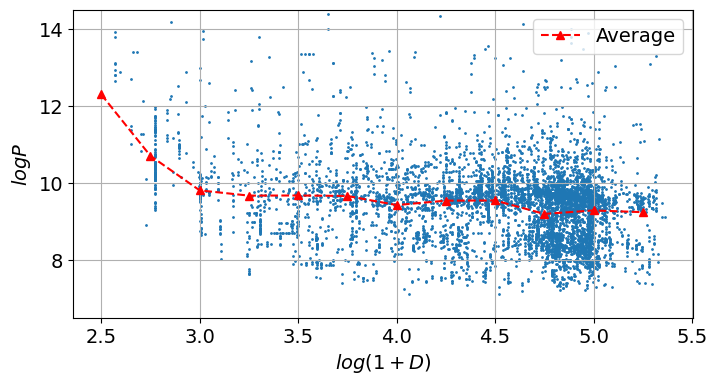}
    \caption{The observations of listing transactions on the Decentraland marketplace in 2022.
    The blue points and red points show individual listing transactions and the average price for every 0.25 intervals in log distance, respectively.}
    \label{fig:price_dist_ex}
\end{figure}

\paragraph{Price dependence on the distance in the close region}
Figure~\ref{fig:price_dist_ex} shows the observations of listing transactions on the Decentraland in 2022.
Prices for parcels close to the center ($\log(1 + D) < 3 \Leftrightarrow D < 20$) are more expensive than those in remote areas, so there appears to be a price discrepancy.
Both listing and sales prices demonstrate similar trends.
Hence, we have to consider a control variable ($I_{close}$) to mitigate the influence of the discrepancy.

\paragraph{Detailed regression result for land parcel prices}
This paragraph summarizes the supplementary information about our regression analysis.
\dkawai{
We apply Eqn. (2) for each one-year period between 2019 and 2023 to accurately identify outliers in the process described below.
In the regression analysis, we found that a small fraction of observations have unrealistically high or low prices, typically several orders of magnitude different from median sales/listing prices.
To exclude such noisy observations, we excluded the observations whose studentized residuals are more than 3.00 as outliers \cite{Weisberg_2005}.
}

Table~\ref{tab:reg_result} summarizes the result of our 
regression analysis.
We perform estimations using the ordinary least square method (OLS).

\paragraph{Generalizability of our results}
\dkawai{A critical question to our regression result is whether the decrease in the negative dependence on the distance in 2021 and 2022 is generalizable to other platforms.
To look into the question, we consider The Sandbox, another major metaverse platform that adopts NFT as a core foundation of land parcels, publishing them as an NFT collection. 
A stark difference between Decentraland and The Sandbox, though, is that the parcels in The Sandbox are gateways to separate metaverse worlds.
Instead of a single ``flat'' metaverse world in Decentraland, users move their mice over parcels on the map to go to metaverse worlds whose links show up once they click parcels in The Sandbox.
This mechanism is different enough from physical real estate that it is unclear whether real-estate pricing theories, such as bid-rent theory, are relevant.
}

\begin{figure}[htbp]
    \centering
    \includegraphics[width=0.85\textwidth]{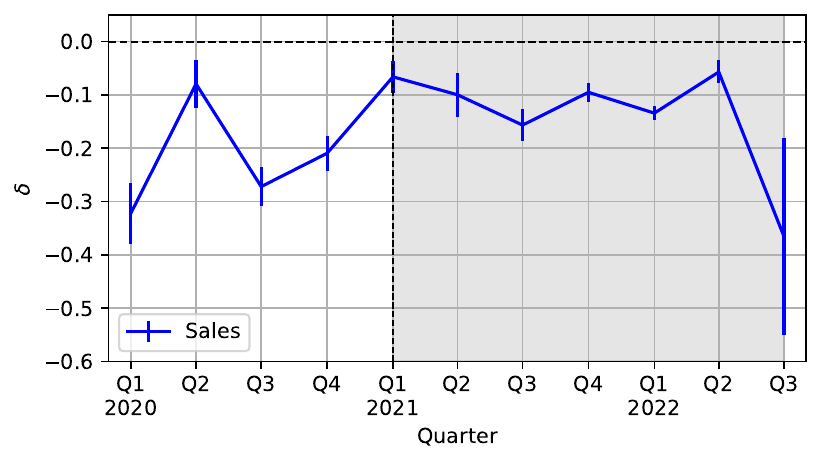}
    \caption{The spatial dependence of land prices ($\delta$) for quarters in 2020-2022.
    The error bars represent 95\% confidence intervals (CIs).
    The grey-colored area shows the 18-month interval surrounding the ``peak hype'' of 2021 Q4.}
    \label{fig:spatial_dependence_sandbox}
\end{figure}

\dkawai{
However, even with the limitation, parcels close to the center of The Sandbox are advantageous in customer exposure since they are displayed when users visit The Sandbox. Hence, we measure how sales prices of The Sandbox's land parcels are affected by their distance from the center. In our analysis, we focus on transactions performed using the ``\textit{atomicMatch\_}'' function in OpenSea's contract.\footnote{\url{https://etherscan.io/address/0x7Be8076f4EA4A4AD08075C2508e481d6C946D12b}}
This is because the majority of sales in 2021 and 2022 were carried out on OpenSea, and 89.4\% of them used the function.
Due to the migration of The Sandbox from Ethereum to the Polygon blockchain,\footnote{\url{https://www.sandbox.game/en/blog/the-sandbox-is-deploying-on-polygon/3060/}} which was announced in June 2022, the data is limited to the period of Q1 2020 to Q3 2022.
We use a simple regression formula,
}
\begin{equation}
    \label{eqn:reg_formula_sandbox}
    \log P_{i,n} = c + \sum_t \left[\alpha_t + \delta_t \log \left(1 + D_i \right) \right] Quarter_{t,n} \, ,
\end{equation}
\dkawai{
where we take the geographical coordinate data from The Sandbox's publicly available API.
The definition of terms in Eqn.~(\ref{eqn:reg_formula_sandbox}) is the same as in Eqn.~(\ref{eq:reg_formula}).
Figure~\ref{fig:spatial_dependence_sandbox} shows the dependence of sales prices ($\log P$) on the distance from the center. It shows that geographical distance matters even less than outside of the price bubble in Decentraland. 
This result hints that our finding for Decentraland can be applied to other metaverse platforms as well.
}

\paragraph{Discussion about saturation hypothesis}
This appendix examines the spatial distribution of developed parcels to determine whether the saturation hypothesis accounts for the surge in prices in late 2021.
An important consideration in distribution analysis is the relationship between the number of parcels and distance. 
It is crucial to note that the number of parcels per distance increases proportionally with distance. 
Comparing the number of developed parcels without considering this factor will lead to an overestimation of development in remote areas.
Therefore, we normalize	the number of developed	parcels within a distance by the total number of parcels in that distance.

\begin{figure}[htbp]
    \centering
    \includegraphics[width=0.85\textwidth]{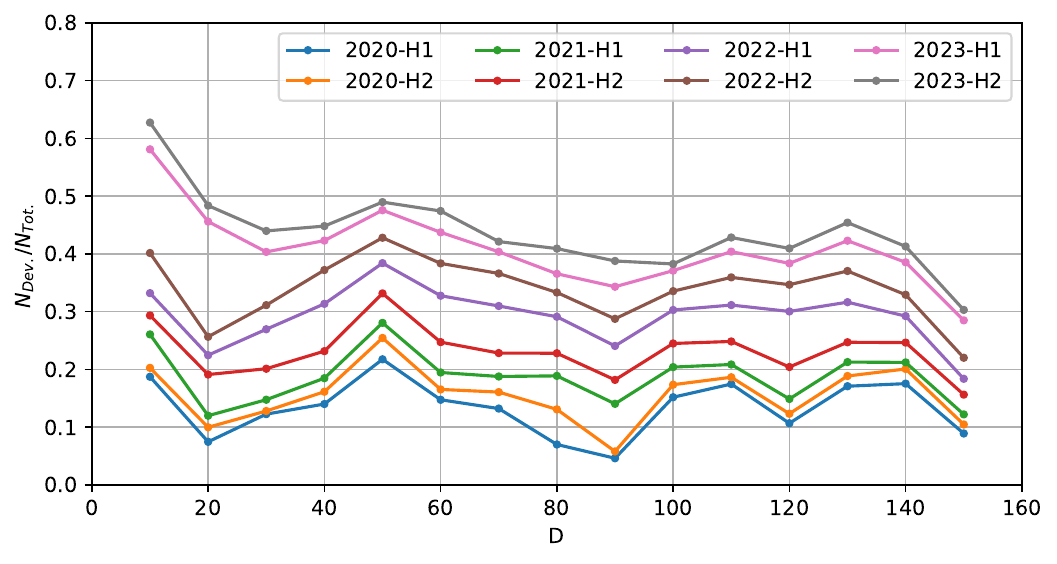}
    \caption{The spatial distribution of developed parcels.
    Lines show the number of developed parcels ($N_{Dev}$) normalized by the total number of parcels ($N_{Tot}$) for every 10-distance interval over half-year periods from 2020 to 2023.
    \label{fig:decentraland_spatial_dist}
    }
\end{figure}

Figure~\ref{fig:decentraland_spatial_dist} shows the number of developed parcels ($N_{Dev}$) normalized by the total number of parcels ($N_{Tot}$) for every 10-distance interval over half-year periods from 2020 to 2023.\footnote{Due to the assumption that developed parcels are never cleared (see Sec.~\ref{subsec:commercial_purpose}), the number of developed parcels increases monotonically over time. It shows the maximum number of parcels developed by the time}
It shows that only 20-30\% of parcels in the map were developed by 2021 Q4, suggesting that the CBD of Decentraland is far from saturation.

\begin{landscape}
\begin{table}[htbp]
    \centering
    \caption{
    \label{tab:reg_result}
    Regression results. The values in parentheses represent the standard errors for estimates.}
    \begin{adjustbox}{width=1.5\textwidth,center}
    \begin{threeparttable}
    \begin{tabular}{llcccclcccclcccclcccclcccc} \toprule
         & &
         \multicolumn{4}{c}{2019} & & 
         \multicolumn{4}{c}{2020} & & 
         \multicolumn{4}{c}{2021} & &
         \multicolumn{4}{c}{2022} & &
         \multicolumn{4}{c}{2023}\\ \cmidrule{3-6} \cmidrule{8-11} \cmidrule{13-16} \cmidrule{18-21} \cmidrule{23-26}
         & &
         Q1 &  Q2 & Q3 & Q4 & & 
         Q1 &  Q2 & Q3 & Q4 & & 
         Q1 &  Q2 & Q3 & Q4 & &
         Q1 &  Q2 & Q3 & Q4 & &
         Q1 &  Q2 & Q3 & Q4 \\ \midrule 
         \multicolumn{2}{l}{\textbf{Listing Price}} \\
         & c &
         \multicolumn{4}{c}{\begin{tabular}{c} 8.767 \\(0.090)\end{tabular}} & &
         \multicolumn{4}{c}{\begin{tabular}{c} 8.676 \\(0.086)\end{tabular}} & &
         \multicolumn{4}{c}{\begin{tabular}{c} 10.226 \\(0.130)\end{tabular}} & &
         \multicolumn{4}{c}{\begin{tabular}{c} 11.378 \\(0.134)\end{tabular}} & &
         \multicolumn{4}{c}{\begin{tabular}{c} 9.036 \\(0.198)\end{tabular}}\\
         & $\alpha_t$ &
         -- &
         \begin{tabular}{c} 0.083\\(0.122)\end{tabular} &
         \begin{tabular}{c} -0.906\\(0.159)\end{tabular} &
         \begin{tabular}{c} 0.493\\(0.163)\end{tabular} & &
         -- &
         \begin{tabular}{c} -0.672\\(0.137)\end{tabular} &
         \begin{tabular}{c} 0.779\\(0.146)\end{tabular} &
         \begin{tabular}{c} 0.371\\(0.157)\end{tabular} & &
         -- &
         \begin{tabular}{c} -0.591\\(0.174)\end{tabular} &
         \begin{tabular}{c} -0.189\\(0.217)\end{tabular} &
         \begin{tabular}{c} 0.704\\(0.169)\end{tabular} & &
         -- &
         \begin{tabular}{c} -1.015\\(0.182)\end{tabular} &
         \begin{tabular}{c} -1.725\\(0.215)\end{tabular} &
         \begin{tabular}{c} -2.518\\(0.256)\end{tabular} & &
         -- &
         \begin{tabular}{c} -0.992\\(0.345)\end{tabular} &
         \begin{tabular}{c} -1.375\\(0.346)\end{tabular} &
         \begin{tabular}{c} -0.269\\(0.387)\end{tabular} \\
         & $\delta_t$ &
         \begin{tabular}{c} -0.449\\(0.019)\end{tabular} &
         \begin{tabular}{c} -0.430\\(0.020)\end{tabular} &
         \begin{tabular}{c} -0.298\\(0.031)\end{tabular} &
         \begin{tabular}{c} -0.423\\(0.031)\end{tabular} & &
         \begin{tabular}{c} -0.385\\(0.018)\end{tabular} &
         \begin{tabular}{c} -0.302\\(0.026)\end{tabular} &
         \begin{tabular}{c} -0.516\\(0.028)\end{tabular} &
         \begin{tabular}{c} -0.436\\(0.030)\end{tabular} & &
         \begin{tabular}{c} -0.388\\(0.029)\end{tabular} &
         \begin{tabular}{c} -0.166\\(0.029)\end{tabular} &
         \begin{tabular}{c} -0.310\\(0.040)\end{tabular} &
         \begin{tabular}{c} -0.257\\(0.026)\end{tabular} & &
         \begin{tabular}{c} -0.359\\(0.029)\end{tabular} &
         \begin{tabular}{c} -0.310\\(0.032)\end{tabular} &
         \begin{tabular}{c} -0.249\\(0.042)\end{tabular} &
         \begin{tabular}{c} -0.188\\(0.051)\end{tabular} & &
         \begin{tabular}{c} -0.283\\(0.043)\end{tabular} &
         \begin{tabular}{c} -0.142\\(0.065)\end{tabular} &
         \begin{tabular}{c} -0.133\\(0.068)\end{tabular} &
         \begin{tabular}{c} -0.352\\(0.076)\end{tabular} \\
         & $I_{District}$ &
         \multicolumn{4}{c}{\begin{tabular}{c} 0.160 \\(0.013)\end{tabular}} & &
         \multicolumn{4}{c}{\begin{tabular}{c} 0.222 \\(0.012)\end{tabular}} & &
         \multicolumn{4}{c}{\begin{tabular}{c} 0.178 \\(0.017)\end{tabular}} & &
         \multicolumn{4}{c}{\begin{tabular}{c} 0.143 \\(0.021)\end{tabular}} & &
         \multicolumn{4}{c}{\begin{tabular}{c} 0.153 \\(0.033)\end{tabular}} \\
         & $I_{Plaza}$ &
         \multicolumn{4}{c}{\begin{tabular}{c} 0.311 \\(0.017)\end{tabular}} & &
         \multicolumn{4}{c}{\begin{tabular}{c} 0.247 \\(0.017)\end{tabular}} & &
         \multicolumn{4}{c}{\begin{tabular}{c} 0.208 \\(0.024)\end{tabular}} & &
         \multicolumn{4}{c}{\begin{tabular}{c} 0.105 \\(0.028)\end{tabular}} & &
         \multicolumn{4}{c}{\begin{tabular}{c} -0.030 \\(0.045)\end{tabular}}\\
         & $I_{Road}$ &
         \multicolumn{4}{c}{\begin{tabular}{c} 0.356 \\(0.019)\end{tabular}} & &
         \multicolumn{4}{c}{\begin{tabular}{c} 0.178 \\(0.018)\end{tabular}} & &
         \multicolumn{4}{c}{\begin{tabular}{c} 0.221 \\(0.023)\end{tabular}} & &
         \multicolumn{4}{c}{\begin{tabular}{c} 0.188 \\(0.028)\end{tabular}} & &
         \multicolumn{4}{c}{\begin{tabular}{c} 0.451 \\(0.044)\end{tabular}} \\
         & $I_{Close}$ &
         \multicolumn{4}{c}{\begin{tabular}{c} 0.375 \\(0.049)\end{tabular}} & &
         \multicolumn{4}{c}{\begin{tabular}{c} 0.747 \\(0.063)\end{tabular}} & &
         \multicolumn{4}{c}{\begin{tabular}{c} 1.066 \\(0.065)\end{tabular}} & &
         \multicolumn{4}{c}{\begin{tabular}{c} 0.899 \\(0.067)\end{tabular}} & &
         \multicolumn{4}{c}{\begin{tabular}{c} 1.522 \\(0.128)\end{tabular}}
         \vspace{5pt} \\
         & Adjusted $R^2$ &
         \multicolumn{4}{c}{0.181} & &
         \multicolumn{4}{c}{0.167} & &
         \multicolumn{4}{c}{0.318} & &
         \multicolumn{4}{c}{0.440} & &
         \multicolumn{4}{c}{0.274}\\
         & Num. of obs. &
         \multicolumn{4}{c}{19,876} & &
         \multicolumn{4}{c}{15,579} & &
         \multicolumn{4}{c}{12,073} & &
         \multicolumn{4}{c}{6,288} & &
         \multicolumn{4}{c}{2,132}\\
         & Log-likelihood &
         \multicolumn{4}{c}{-24,379} & &
         \multicolumn{4}{c}{-17,468} & &
         \multicolumn{4}{c}{-15,716} & &
         \multicolumn{4}{c}{-7,239} & &
         \multicolumn{4}{c}{-2,321}\\
         & AIC &
         \multicolumn{4}{c}{$4.878\times 10^4$} & &
         \multicolumn{4}{c}{$3.496\times 10^4$} & &
         \multicolumn{4}{c}{$3.146\times 10^4$} & &
         \multicolumn{4}{c}{$1.450\times 10^4$} & &
         \multicolumn{4}{c}{$4.67\times 10^3$}\\
         & BIC &
         \multicolumn{4}{c}{$4.89\times 10^4$} & &
         \multicolumn{4}{c}{$3.51\times 10^4$} & &
         \multicolumn{4}{c}{$3.15\times 10^4$} & &
         \multicolumn{4}{c}{$1.46\times 10^4$} & &
         \multicolumn{4}{c}{$4.74\times 10^3$}\\
         & Prob(F-stat.) &
         \multicolumn{4}{c}{0.00} & &
         \multicolumn{4}{c}{0.00} & &
         \multicolumn{4}{c}{0.00} & &
         \multicolumn{4}{c}{0.00} & &
         \multicolumn{4}{c}{0.00} \vspace{5pt}\\
         \multicolumn{2}{l}{\textbf{Sales Price}} \\
         & c &
         \multicolumn{4}{c}{\begin{tabular}{c} 7.925 \\(0.174)\end{tabular}} & &
         \multicolumn{4}{c}{\begin{tabular}{c} 7.814 \\(0.172)\end{tabular}} & &
         \multicolumn{4}{c}{\begin{tabular}{c} 10.027 \\(0.151)\end{tabular}} & &
         \multicolumn{4}{c}{\begin{tabular}{c} 9.976 \\(0.113)\end{tabular}} & &
         \multicolumn{3}{c}{\begin{tabular}{c} 8.347 \\(0.328)\end{tabular}}\\
         & $\alpha_t$ &
         -- &
         \begin{tabular}{c} -0.861\\(0.274)\end{tabular} &
         \begin{tabular}{c} 0.146\\(0.292)\end{tabular} &
         \begin{tabular}{c} -0.966\\(0.302)\end{tabular} & &
         -- &
         \begin{tabular}{c} -0.411\\(0.334)\end{tabular} &
         \begin{tabular}{c} 0.286\\(0.245)\end{tabular} &
         \begin{tabular}{c} 0.178\\(0.320)\end{tabular} & &
         -- &
         \begin{tabular}{c} -1.060\\(0.223)\end{tabular} &
         \begin{tabular}{c} -0.852\\(0.257)\end{tabular} &
         \begin{tabular}{c} -0.424\\(0.188)\end{tabular} & &
         -- &
         \begin{tabular}{c} -0.503\\(0.198)\end{tabular} &
         \begin{tabular}{c} -0.434\\(0.296)\end{tabular} &
         \begin{tabular}{c} -1.795\\(0.261)\end{tabular} & &
         -- &
         \begin{tabular}{c} -0.067\\(0.503)\end{tabular} &
         \begin{tabular}{c} -1.615\\(0.571)\end{tabular} &
         \begin{tabular}{c} -0.671\\(0.510)\end{tabular}  \\
         & $\delta_t$ &
         \begin{tabular}{c} -0.363\\(0.037)\end{tabular} &
         \begin{tabular}{c} -0.165\\(0.047)\end{tabular} &
         \begin{tabular}{c} -0.428\\(0.054)\end{tabular} &
         \begin{tabular}{c} -0.237\\(0.055)\end{tabular} & &
         \begin{tabular}{c} -0.260\\(0.036)\end{tabular} &
         \begin{tabular}{c} -0.239\\(0.062)\end{tabular} &
         \begin{tabular}{c} -0.307\\(0.041)\end{tabular} &
         \begin{tabular}{c} -0.293\\(0.058)\end{tabular} & &
         \begin{tabular}{c} -0.431\\(0.033)\end{tabular} &
         \begin{tabular}{c} -0.097\\(0.038)\end{tabular} &
         \begin{tabular}{c} -0.170\\(0.046)\end{tabular} &
         \begin{tabular}{c} -0.029\\(0.026)\end{tabular} & &
         \begin{tabular}{c} -0.103\\(0.024)\end{tabular} &
         \begin{tabular}{c} -0.182\\(0.037)\end{tabular} &
         \begin{tabular}{c} -0.287\\(0.061)\end{tabular} &
         \begin{tabular}{c} -0.111\\(0.053)\end{tabular} & &
         \begin{tabular}{c} -0.182\\(0.069)\end{tabular} &
         \begin{tabular}{c} -0.231\\(0.086)\end{tabular} &
         \begin{tabular}{c} 0.016\\(0.109)\end{tabular} &
         \begin{tabular}{c} -0.204\\(0.042)\end{tabular}\\
         & $I_{District}$ &
         \multicolumn{4}{c}{\begin{tabular}{c} 0.103 \\(0.023)\end{tabular}} & &
         \multicolumn{4}{c}{\begin{tabular}{c} 0.163 \\(0.023)\end{tabular}} & &
         \multicolumn{4}{c}{\begin{tabular}{c} 0.117 \\(0.019)\end{tabular}} & &
         \multicolumn{4}{c}{\begin{tabular}{c} 0.046 \\(0.018)\end{tabular}} & &
         \multicolumn{4}{c}{\begin{tabular}{c} -0.034 \\(0.042)\end{tabular}} \\
         & $I_{Plaza}$ &
         \multicolumn{4}{c}{\begin{tabular}{c} 0.136 \\(0.038)\end{tabular}} & &
         \multicolumn{4}{c}{\begin{tabular}{c} 0.129 \\(0.037)\end{tabular}} & &
         \multicolumn{4}{c}{\begin{tabular}{c} 0.105 \\(0.027)\end{tabular}} & &
         \multicolumn{4}{c}{\begin{tabular}{c} 0.152 \\(0.026)\end{tabular}} & &
         \multicolumn{4}{c}{\begin{tabular}{c} 0.118 \\(0.060)\end{tabular}}\\
         & $I_{Road}$ &
         \multicolumn{4}{c}{\begin{tabular}{c} 0.280 \\(0.028)\end{tabular}} & &
         \multicolumn{4}{c}{\begin{tabular}{c} 0.029 \\(0.028)\end{tabular}} & &
         \multicolumn{4}{c}{\begin{tabular}{c} 0.115 \\(0.022)\end{tabular}} & &
         \multicolumn{4}{c}{\begin{tabular}{c} 0.151 \\(0.023)\end{tabular}} & &
         \multicolumn{4}{c}{\begin{tabular}{c} 0.186 \\(0.055)\end{tabular}} \\
         & $I_{Close}$ &
         \multicolumn{4}{c}{\begin{tabular}{c} 0.697 \\(0.107)\end{tabular}} & &
         \multicolumn{4}{c}{\begin{tabular}{c} 0.725 \\(0.179)\end{tabular}} & &
         \multicolumn{4}{c}{\begin{tabular}{c} 0.635 \\(0.104)\end{tabular}} & &
         \multicolumn{4}{c}{\begin{tabular}{c} 0.270 \\(0.110)\end{tabular}} & &
         \multicolumn{4}{c}{\begin{tabular}{c} 1.332 \\(0.277)\end{tabular}}
         \vspace{5pt} \\
         & Adjusted $R^2$ &
         \multicolumn{4}{c}{0.267} & &
         \multicolumn{4}{c}{0.169} & &
         \multicolumn{4}{c}{0.571} & &
         \multicolumn{4}{c}{0.770} & &
         \multicolumn{4}{c}{0.520}\\
         & Num. of obs. &
         \multicolumn{4}{c}{2,151} & &
         \multicolumn{4}{c}{1,877} & &
         \multicolumn{4}{c}{3,708} & &
         \multicolumn{4}{c}{1,712} & &
         \multicolumn{4}{c}{305}\\
         & Log-likelihood &
         \multicolumn{4}{c}{-1,567} & &
         \multicolumn{4}{c}{-1,265} & &
         \multicolumn{4}{c}{-2,838} & &
         \multicolumn{4}{c}{-602} & &
         \multicolumn{4}{c}{-103}\\
         & AIC &
         \multicolumn{4}{c}{3,157} & &
         \multicolumn{4}{c}{2,554} & &
         \multicolumn{4}{c}{5,699} & &
         \multicolumn{4}{c}{1,228} & &
         \multicolumn{4}{c}{230}\\
         & BIC &
         \multicolumn{4}{c}{3,225} & &
         \multicolumn{4}{c}{2,621} & &
         \multicolumn{4}{c}{5,774} & &
         \multicolumn{4}{c}{1,293} & &
         \multicolumn{4}{c}{275}\\
         & Prob(F-stat.) &
         \multicolumn{4}{c}{0.00} & &
         \multicolumn{4}{c}{0.00} & &
         \multicolumn{4}{c}{0.00} & &
         \multicolumn{4}{c}{0.00} & &
         \multicolumn{4}{c}{0.00}\\
         \bottomrule
    \end{tabular}
    \end{threeparttable}
  \end{adjustbox}
\end{table}
\end{landscape}

\end{document}